\documentclass[epj,final]{svjour}
\usepackage{graphicx}
\usepackage{amsmath}
\begin{document}

\title{Correction algorithm for finite sample statistics}

\author{Thorsten P\"oschel\inst{1} \and Werner Ebeling\inst{1} \and
  Cornelius Fr\"ommel\inst{1} \and Rosa Ram\'{\i}rez\inst{2}}
\authorrunning{Thorsten P\"oschel et al.}  \institute{ \inst{1}
  Humboldt Universit\"at zu Berlin, Charit\'e, Institut f\"ur
  Biochemie, Monbijoustra{\ss}e 2, D-10117 Berlin,
  Germany\\
  \inst{2} Centre Europ\'een de Calcul Atomique et Mol\'eculaire
  (CECAM), Ecole Normale Sup\'erieure, 46, All\'ee d'Italie,\\
  Fr-69007 Lyon, France}

\date{Received: \today /  Revised version: }

\abstract{Assume in a sample of size $M$ one finds $M_i$
  representatives of species $i$ with $i=1\dots N^{*}$. The normalized
  frequency $p^*_i\equiv M_i/M$, based on the finite sample, may
  deviate considerably from the true probabilities $p_i$. We propose a
  method to infer rank-ordered true probabilities $r_i$ from measured
  frequencies $M_i$. We show that the rank-ordered probabilities
  provide important informations on the system, e.g., the true number
  of species, the Shannon- and the Renyi-entropies.}   \maketitle

\section{Introduction}

In experimental work one frequently faces the problem to determine the
probabilities of occurrence (or concentrations) $p_1$, $p_2,\dots,
p_N$ of species 1, 2,\dots, $N$. The probability of species $i$ is
defined by
\begin{eqnarray}
p_i&=&\lim\limits_{M\rightarrow\infty}\frac{M_i}{M}\,,~~~~~~i=1\dots N \label{cdef}\\
M&=&\sum\limits_{j=1}^N M_j \,,
\end{eqnarray}
with $M_i$ being the number of representatives of the species $i$
found in a sample of size $M$. $N$ is the number of different species
which will appear in a sample of {\em infinite} size. Of course $M$
will never be infinite in reality, but a number which is determined
mainly by the experimental effort, i.e., usually costs and time. In
this article the term ``species'' is not used in its strict
phylogenetic sense, but it stands as a synonym for ``distinguishable
events which are members of a statistical ensemble''.

A prominent example where we cannot reliably infer the probabilities
from counted frequencies is the distribution of words of length $n$ in
nucleotide sequences such as DNA. Since we have an alphabet of four
letters, there are $4^n$ words which in principle could be
constructed. Even for moderate values of $n$, the number of words
exceeds the size of {\em any} available data base. If we want to
compute, the entropy of the word distribution in biosequences we have
to apply, therefore, correction methods, e.g.
\cite{grassberger,schmitt0,Schmitt,Herzel,PER,ep,EPA}.

Virtually each experimental measurement of concentrations (or
probabilities) is affected by finite size effects due to the feasible
number $M$ of samples which can be investigated. In a real measurement
one cannot even expect to find the correct number $N$ of species.
Instead, in general, a smaller number $N^*$ is found, depending on the
sample size $M$. We will show that, even if $M$ is a rather large
number, the deviations of the observed relative frequencies
\begin{equation}
p^*_{i}=\frac{M_i}{M}\,, \qquad \mbox{$M$ finite}
\end{equation}
from the true probabilities $p_i$ as defined by Eq. \eqref{cdef} may
be significant. A method to deduce true probabilities from measured
relative frequencies is, therefore, highly desirable.

The aim of this article is to propose a method to correct relative
frequencies $p_i^*=M_i/M$ in a finite sample of size $M$ in a way to
approximate the true probabilities $p_i$ which would be obtained if a
sample of infinite size was investigated.  Our method is based on the
idea, that the estimation of rank-ordered probabilities is by orders
of magnitude more easy than the estimation of the species-ordered
probabilities. This is due to the fact that in the rank-ordering
procedure the exact relation between the species number $i$ and the
probability $p_i$ is ignored. This way it remains to estimate the
shape of a function $r_i$ which is monotonously decreasing with the
rank $i$ (see Sec. \ref{sec:rank} for the definition of $r_i$). The
large interest in rank-ordered distributions is based on the fact that
several characteristic quantities as the Renyi-entropies \cite{Renyi}
\begin{equation}
  \label{eq:Renyi}
H^{(q)} = \frac{1}{1-q}\log \left(\sum p_i^q\right) = 
\frac{1}{1-q}\log \left(\sum r_i^q\right) 
\end{equation}
are invariant with respect to the ordering. Therefore, the
rank-ordered distribution suffices to compute the Renyi entropies. 
We notice the important relations $M = \exp H^{(0)}$, $H = H^{(1)}$.
In other words, the true number of species $M$ as well as the 
Shannon entropies $H$ are exactly calculable from the rank-ordered
distributions $r_i$. In the last section we will show, how our method
can be extended also to estimate any mean value of statistical variables.  

\section{Species ordered distributions and rank-ordered distributions}
\label{sec:rank} 
\begin{figure}[htbp]
  \centerline{\includegraphics[width=7cm,clip]{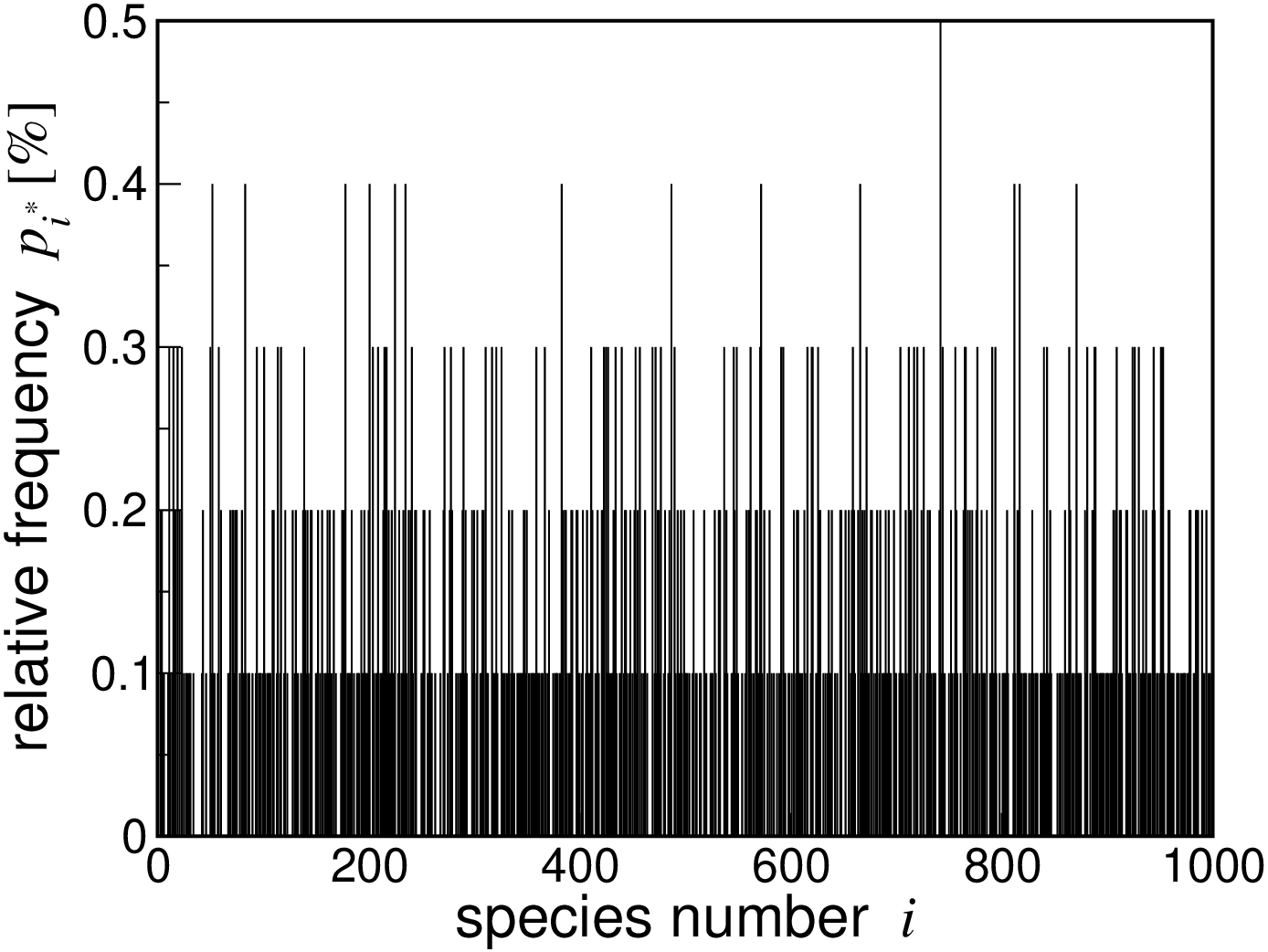}}
  \centerline{\includegraphics[width=7cm,clip]{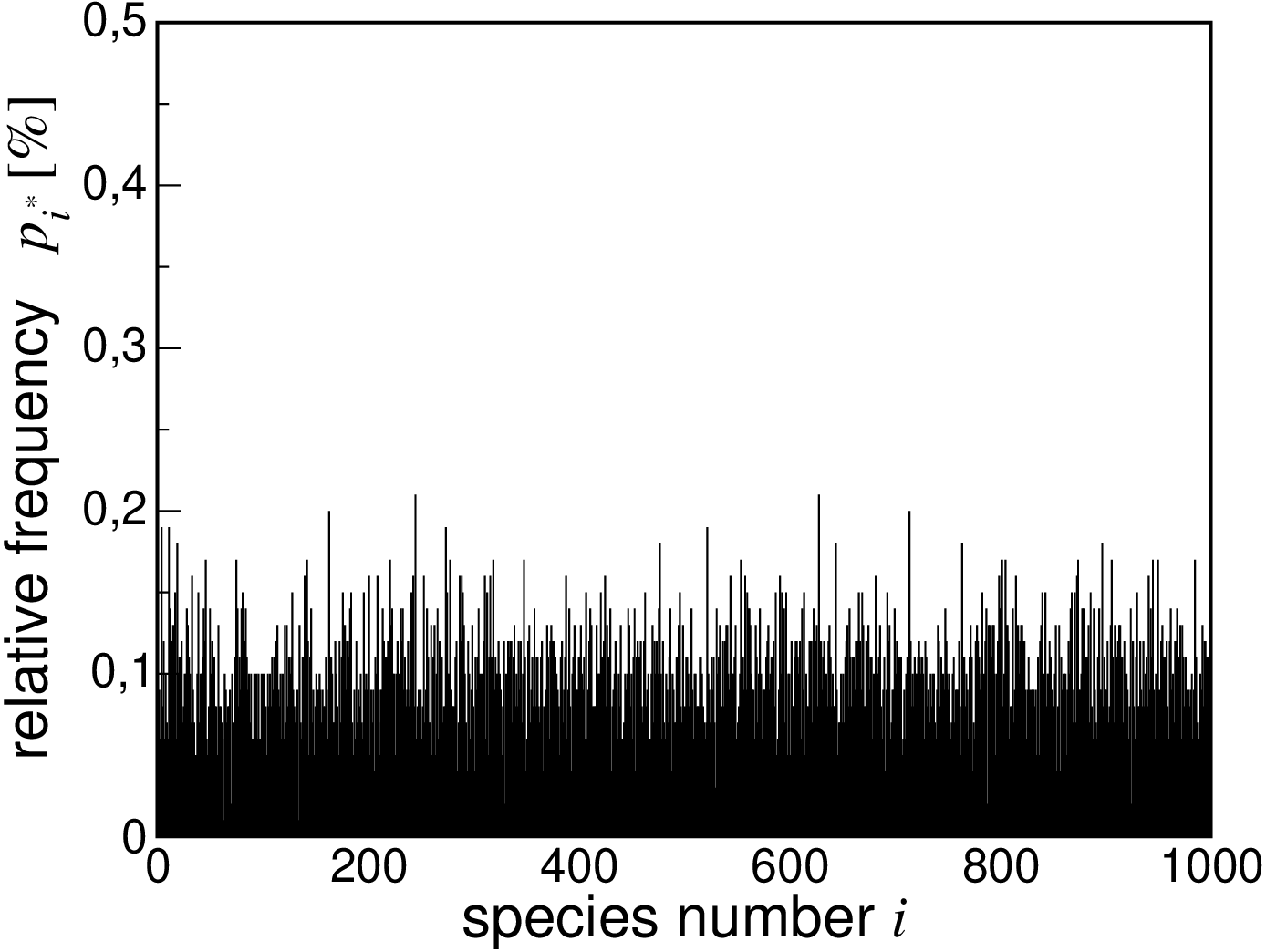}}
  \centerline{\includegraphics[width=7cm,clip]{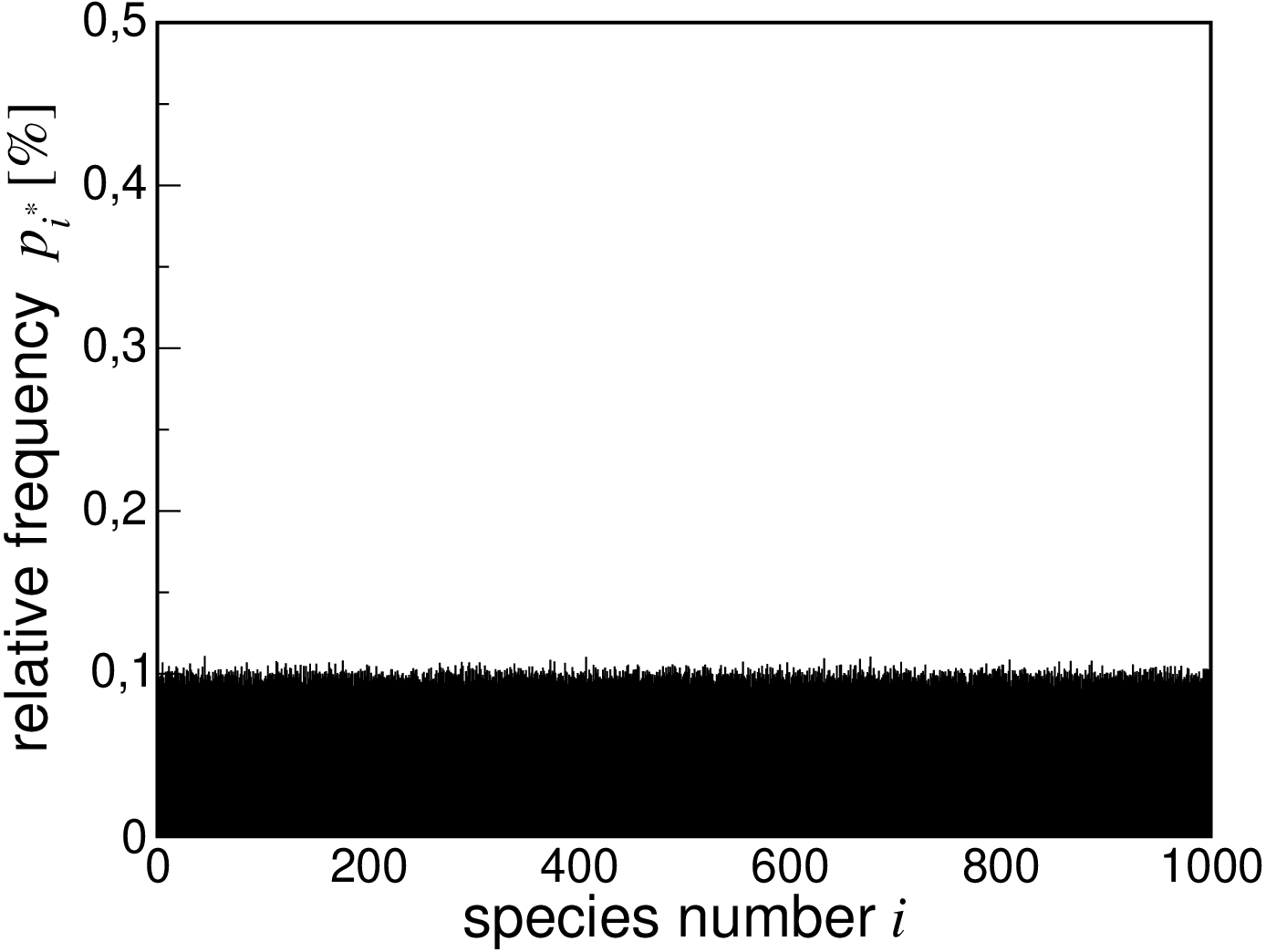}}
  \caption{Observed relative frequencies of $N=1000$ equidistributed 
    species found in samples of size $M=10^3$ (top), $M=10^4$
    (middle), and $M=10^6$ (bottom).}
\label{fig:unzipf}
\end{figure}

Assume we draw a sample from a system of $N$ different species which
are equally distributed $p_1=p_2=\dots=p_N=1/N$. Symbols with upper
index $*$ such as $p_i^*$ denote {\em observed} quantities in a finite
sample of size $M$. Obviously, if $M$ is large enough the relative
frequencies approach the probabilities, $p^*_{1} \rightarrow p_1$,
$p^*_{2}\rightarrow p_2 \dots$, $p^*_{N}\rightarrow p_N$ due to Eq.
\eqref{cdef}. Figure \ref{fig:unzipf} shows the observed relative
frequencies for three different sample sizes $M$ for 1000
equidistributed species with $p_1=p_2=\dots =p_{1000}=1/1000$.
For this figure we produced uniformly distributed random integers from
the interval $[0,999]$ and counted the occurrences of each number. As
expected, with increasing sample size $M$ the distribution resembles
more and more the equidistribution in agreement with the true
probabilities $p_i$. Nevertheless, the deviations of the relative
frequencies from the probabilities are significant: even in the case
of rather large {\em relative sample size} $M/N=1000$ (Fig.
\ref{fig:unzipf}, bottom) the deviation of the relative frequencies
from the probabilities can be as large as $p^*_{j}/p_j\approx 1.11$.
For the case $M/N=1$ (top of the figure) we can see that many species
have not been found at least once.

\begin{figure}[htbp]
  \centerline{\includegraphics[width=7cm,clip]{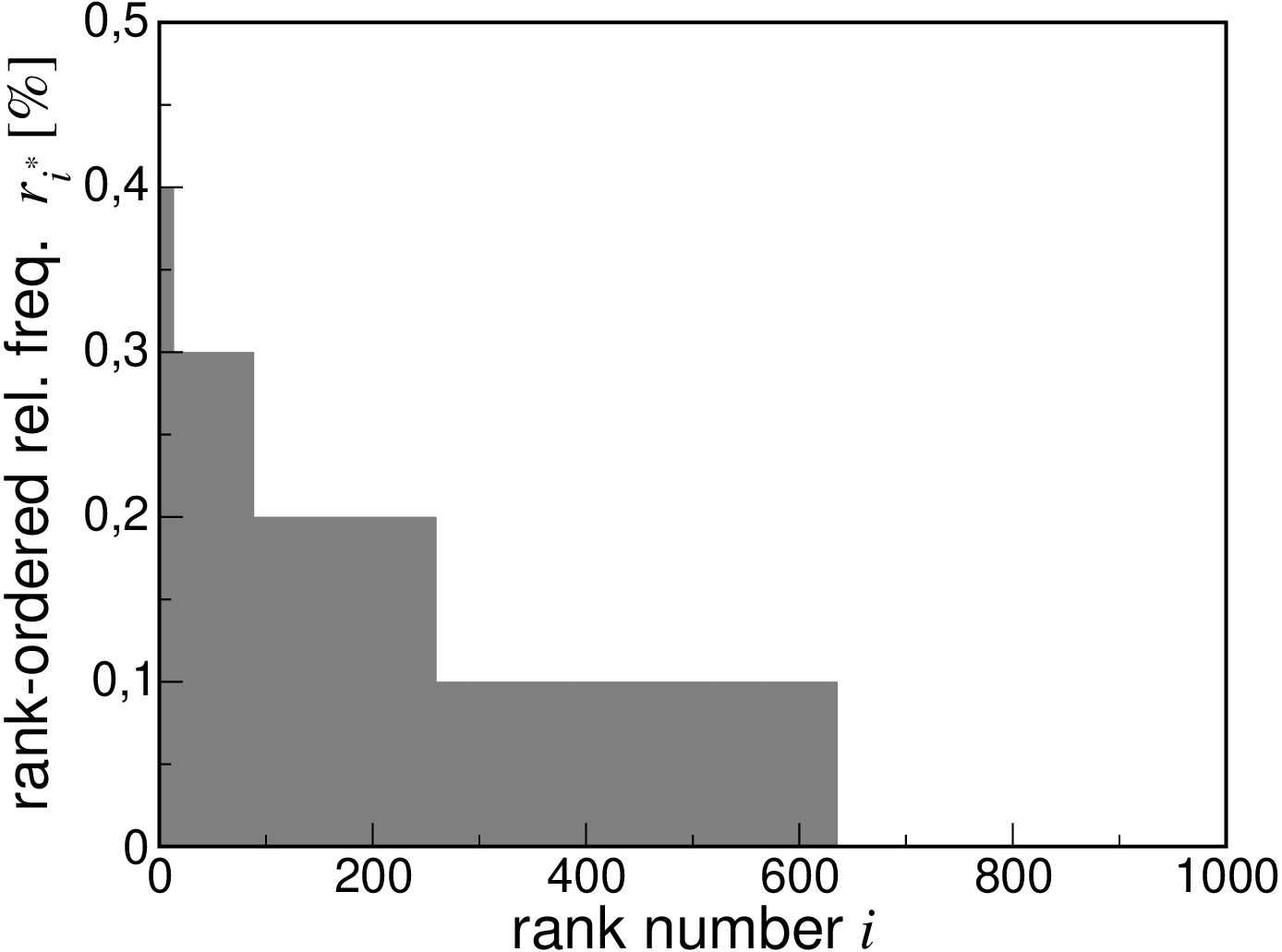}}
  \centerline{\includegraphics[width=7cm,clip]{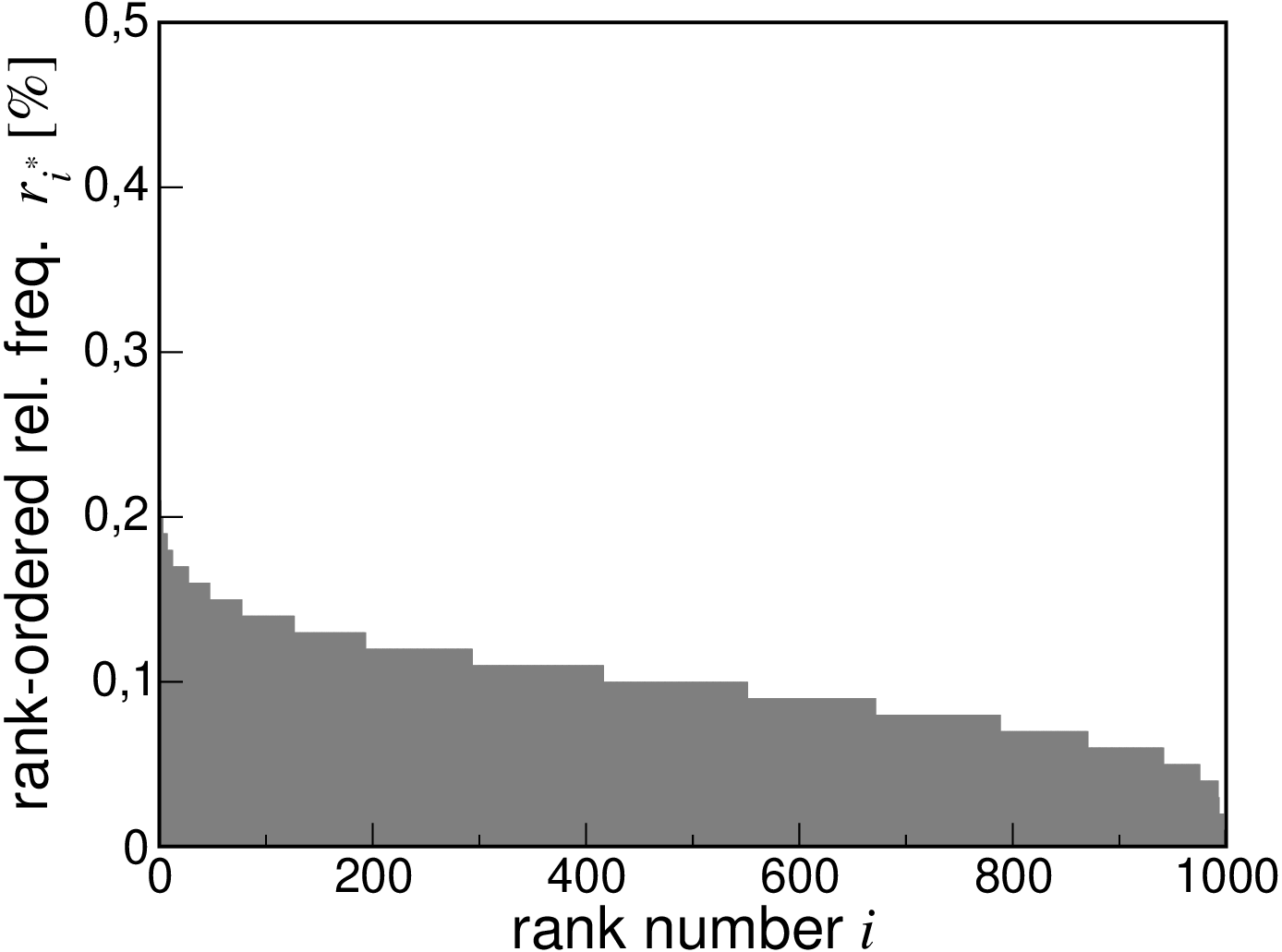}}
\vspace*{-5.0cm}\centerline{~~~~~~~~~~~~~~~~\includegraphics[width=4.5cm,clip]{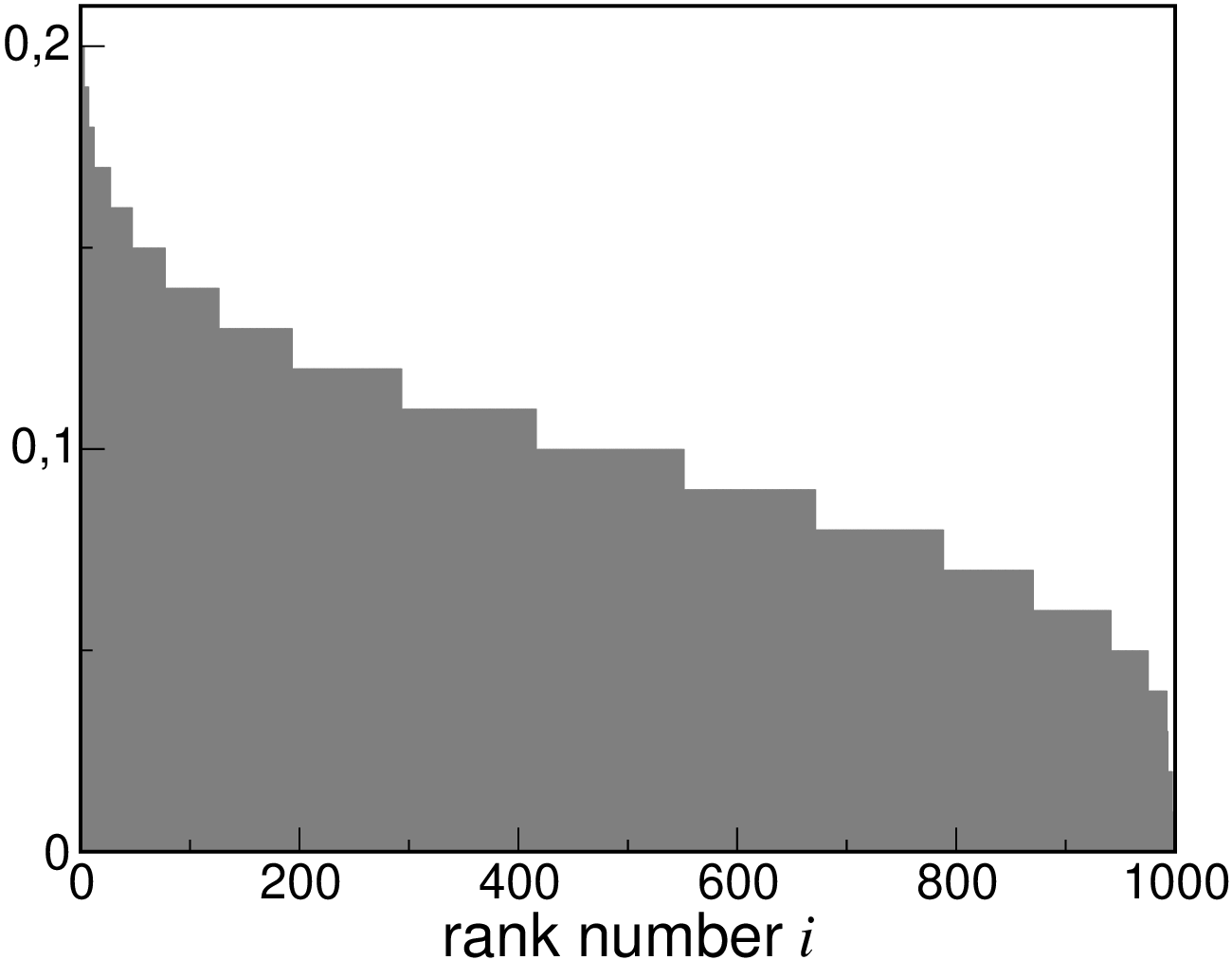}}\vspace*{2cm}
  \centerline{\includegraphics[width=7cm,clip]{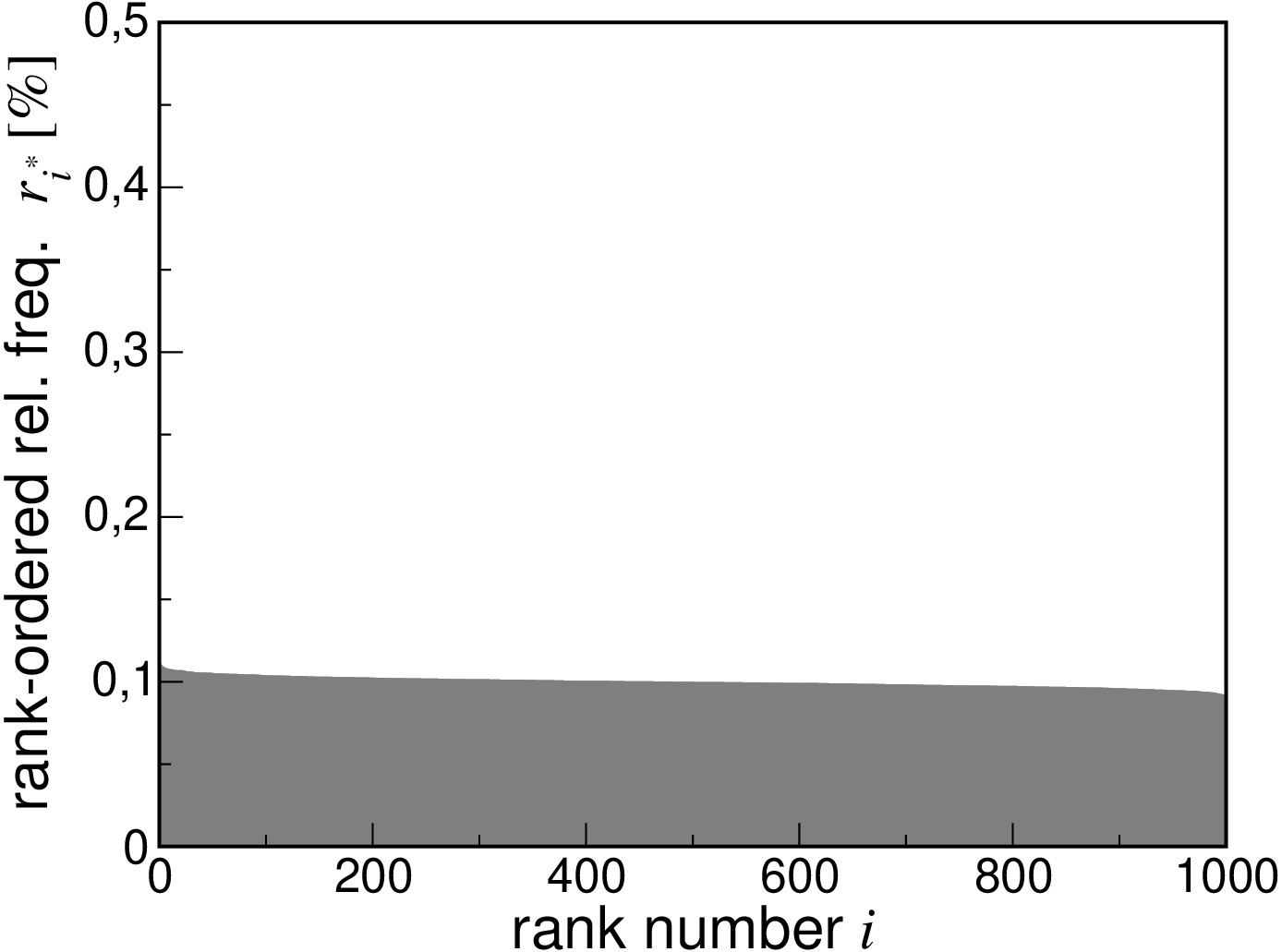}}
\vspace*{-5.0cm}\centerline{~~~~~~~~~~~~~~~~\includegraphics[width=4.5cm,clip]{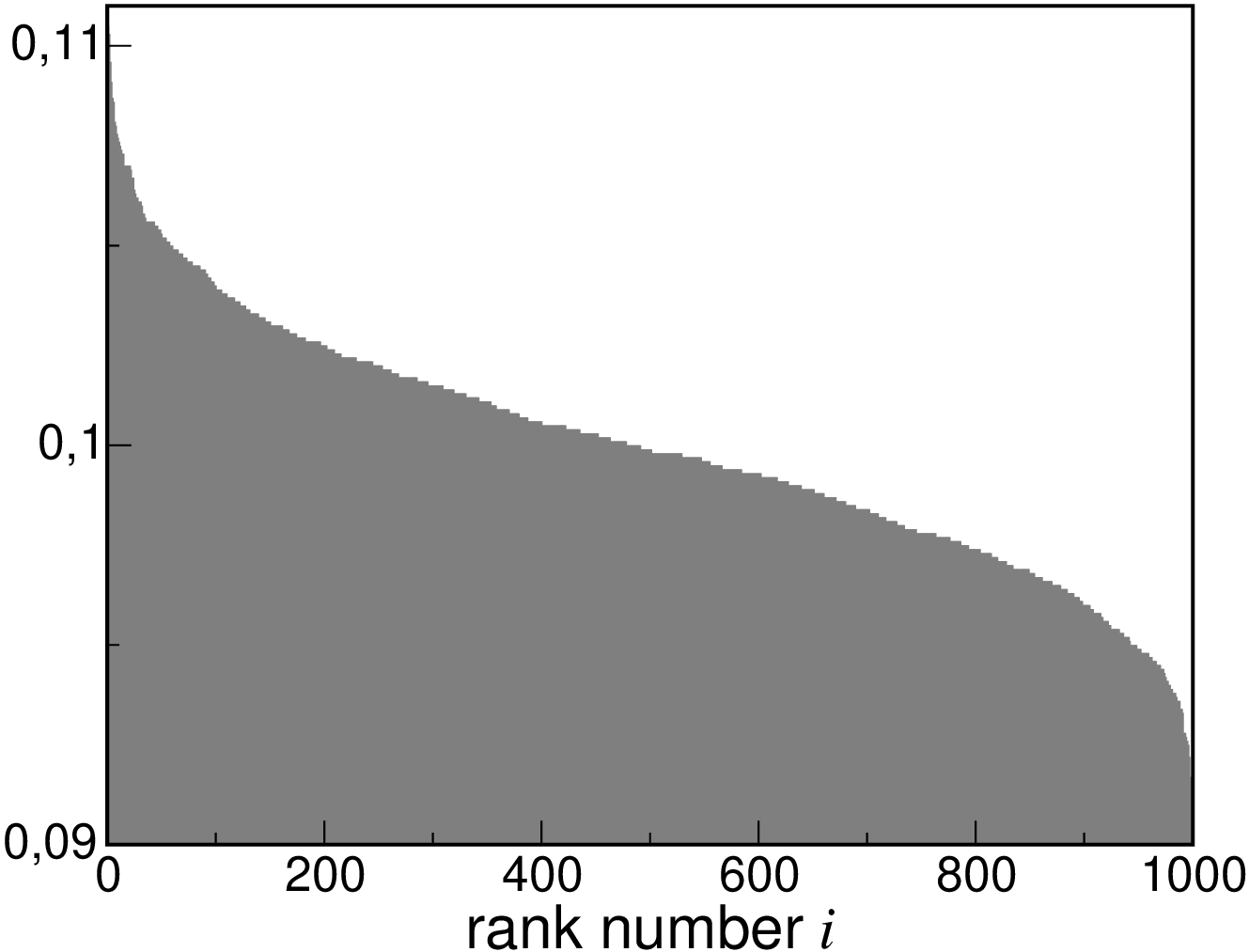}}\vspace*{2cm}
  \caption{Rank-ordered observed relative frequencies $N=1000$ 
    equidistributed species in a sample of size $M=10^3$ (top),
    $M=10^4$ (middle), and $M=10^6$ (bottom). The inserts show the
    same data with higher resolution.}
  \label{fig:zipf}
\end{figure}

To quantify the deviations and for practical purposes that will be
motivated below, we will use the data representation given in Fig.
\ref{fig:zipf}. Here the same data as in Fig. \ref{fig:unzipf} are
displayed, however, the abscissa does not show the species label $i$
but the species are ordered according to the frequency of their
occurrence in the sample. This means the species which occurs with the
largest number of representatives appears at the first position (1) of
the abscissa, the species found with the second largest frequency is
labeled $2$, etc. We call this representation {\em rank-ordered
  distribution of frequencies} where $r^*_i$ is the observed relative
frequency of the species at rank $i$.

Figure \ref{fig:zipf} clearly reveals that even for large relative
sample size $M/N$ the observed rank-ordered frequencies $r^*_i$ may
deviate considerably from the probabilities $p_i$. For smaller sample
size $M/N=1$ (top of the figure) about 1/3 of the species are not even
found once, i.e., the observed number of species may be smaller than
the true number, $N^*\le N$.

The rank-ordered relative frequencies $r^*_i$ form, by definition,
always a decaying function.  In the limit $M\to \infty$ this function
approaches the rank-ordered probabilities $r_i$ which coincide with
$p_i$ for the case of the equidistribution as well as if the
probabilities $p_i$ are decaying with increasing label number $i$ of
the species (see examples in the following sections). As mentioned,
this limit is difficult to achieve when $N$ is large. For $M=10^4$
(Fig. \ref{fig:zipf}, middle) we find a distribution that is far from
being uniform. Even for $M=10^6$ the inset shows a rank-ordered
distribution which deviates considerably from the equidistribution.
Hence, from an observation one might erroneously conclude that the
events are non-equally distributed.

The rank-ordered probabilities $r_i$ ($i$ is the rank index) contain
less information than the species-ordered distribution $p_i$ since the
co-ordination {\em species $\leftrightarrow$ probability} is lost. The
problem to infer the rank-ordered probabilities $r_i$ from a sample of
size $M$ is, therefore, a much simpler problem than to infer the
species related probabilities $p_i$. In general, the rank-ordered
distribution $r_i$ contains about $N!$ times less information than the
species-number ordered distribution $p_i$, since about $N!$
species-ordered distributions correspond the same rank-ordered
distribution:
\begin{equation}
  \label{eq:fak}
  \{r_i,~i=1,\dots,N\} \leftarrow \left\{
    \begin{tabular}{l}
$p_1, p_2, p_3 \dots p_N$\\
$p_2, p_1, p_3 \dots p_N$\\
$p_3, p_2, p_1 \dots p_N$\\
\dots\\
$p_j,~j=1,\dots,N,~\{j\}=\mbox{perm}\{i\}$
    \end{tabular}
\right.
\end{equation}
More precisely, the number of species-number ordered distributions is
slightly smaller than the number of permutations of the species
numbers $N!$ since there might be species which occur at the same
probability so that their permutation does not affect the
distribution.

From these arguments we conclude that it is about $N!$ times simpler
to infer the rank-ordered probabilities $r_i$ from the investigation
of a sample of size $M$ than the species-ordered probabilities $p_i$.
Or, in other words, a sample of size $M$ allows to determine the
rank-ordered probabilities up to a much higher accuracy than the
species-ordered probabilities.

Before coming to the main point, the estimate of probabilities from
finite sample observations, it is helpful first to consider the
inverse problem.

\section{Predicting observed relative frequencies from a probability 
  distribution}
\label{sec:cluster}
\subsection{Equidistributed species}

For the description of the species-ordered observed relative
frequencies $\{p_i^*, i=1,\dots,N\}$, in general $N-1$ numbers are
required, whereas for the corresponding rank-ordered relative
frequencies, $\{r_i^*\}$, it is sufficient to specify, how many
species did not appear in our sample (this quantity will be denoted by
$k_0$), how many species occurred with one representative ($k_1$), how
many with two representatives ($k_2$), etc. An observed rank-ordered
distribution of relative frequencies is, hence, determined by a set of
occupation numbers $\{k_i, i=0,1,2,..,M\}$. In this section we
describe a method to predict the observed rank-ordered relative
frequencies $r^*_i$ from a probability distribution, $p_i$, for finite
sample size $M$.

The observed distribution $\{r^*_i\}$ is characterized by the cluster
distribution $\{k_j\}$: the number of species that appear with $j$
representatives each in a sample of size $M$. We define the
probability distribution $p_c\left(k_i,i\right)$ to find {\em exactly}
$k_i$ species each occurring with precisely $i$ representatives in a
sample of size $M$. With the normalization conditions
\begin{eqnarray}
  \label{eq:condition1}
  \sum\limits_{i=0}^M k_i= N~~&& \mbox{(total number of species)}\\
  \sum\limits_{i=0}^M i\,k_i = M~~&& \mbox{(total number of individuals)}\,,
  \label{eq:condition2}
\end{eqnarray}
for the case of equidistributed species this distribution reads
\cite{JK} (see also \cite{JT,TJ,TPJanf})
\begin{equation}
p_c\left(k_i,i\right)=\frac{M!}{N^M}\sum\limits_{j=k_i}^{\lfloor M/i\rfloor}
(-1)^{(j-k_i)}\left(j\atop k_i\right)
\frac{(N-j)^{(M-ji)}}{\left(i!\right)^j (M-ij)!}\,,
\label{eq:exact}
\end{equation}
where $\lfloor x\rfloor$ denotes the integer of $x$.

The observable $\left<k_i\right>$, i.e., the average number of species
that occur with $i$ representatives when a sample of size $M$ is
drawn, is the first moment of this probability distribution
\cite{JT,TJ} $\left<k_i\right> =\sum_{k_i} k_i \,
p_c\left(k_i,i\right)$, where the summation is to be performed over
all cluster distributions which are in agreement with Eqs.
\eqref{eq:condition1} and \eqref{eq:condition2}:
\begin{equation}
  \left<k_i\right>= \left(M\atop i\right) N^{(1-i)}  \left(1-\frac{1}{N}\right)^{(M-i)}.
\label{eq:Mom}
\end{equation}
The occupation numbers $i=0,1,2,\cdots$ are called the $i$-clusters;
$\left<k_0\right>$ is then the average size of the cluster of species
which do not appear in our sample, $\left<k_1\right>$ is the size of
the cluster of species which appear with one representative, etc.

\begin{figure}[htb]
\centerline{\includegraphics[width=7cm,clip]{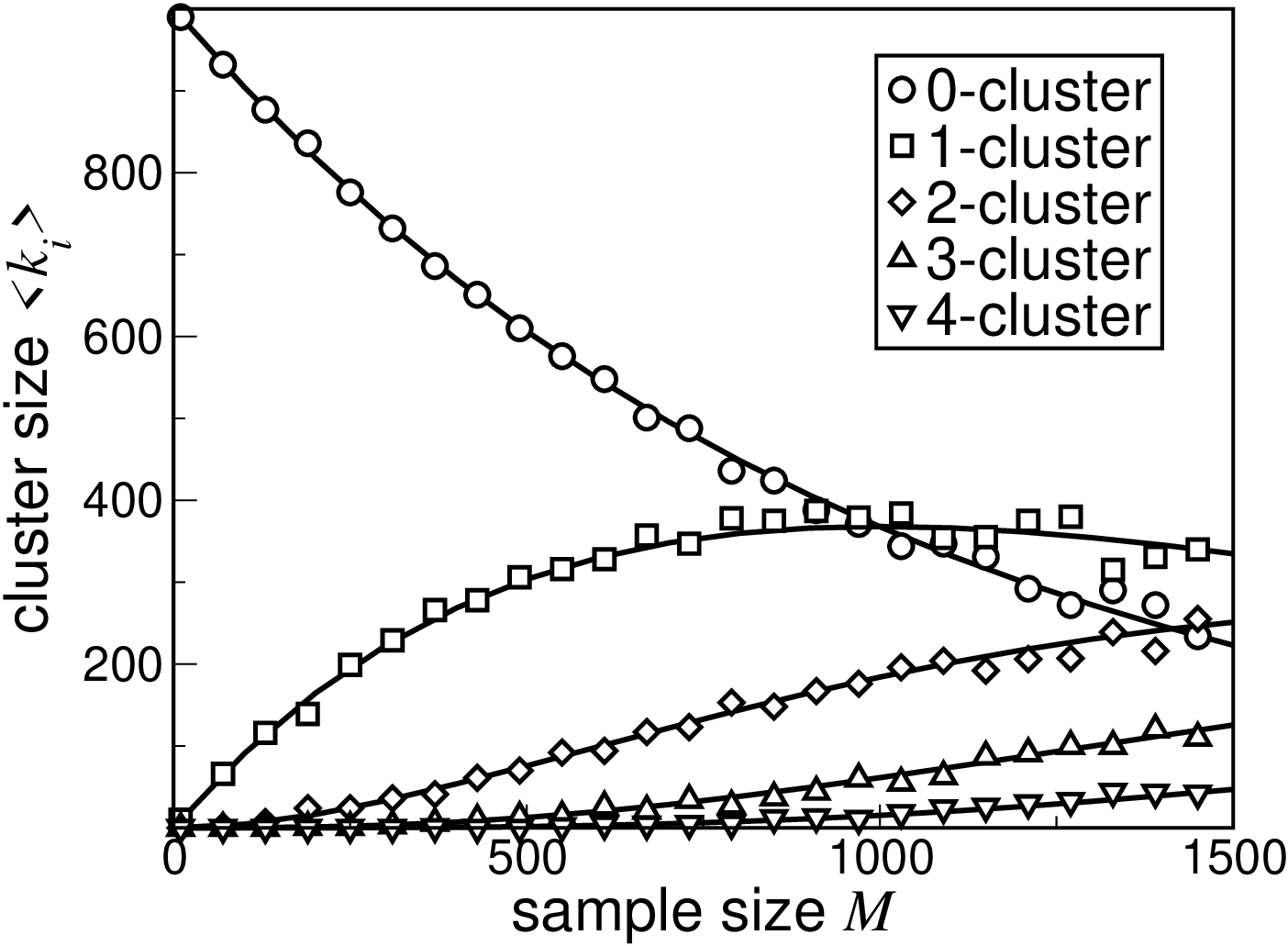}}
\centerline{\includegraphics[width=7cm,clip]{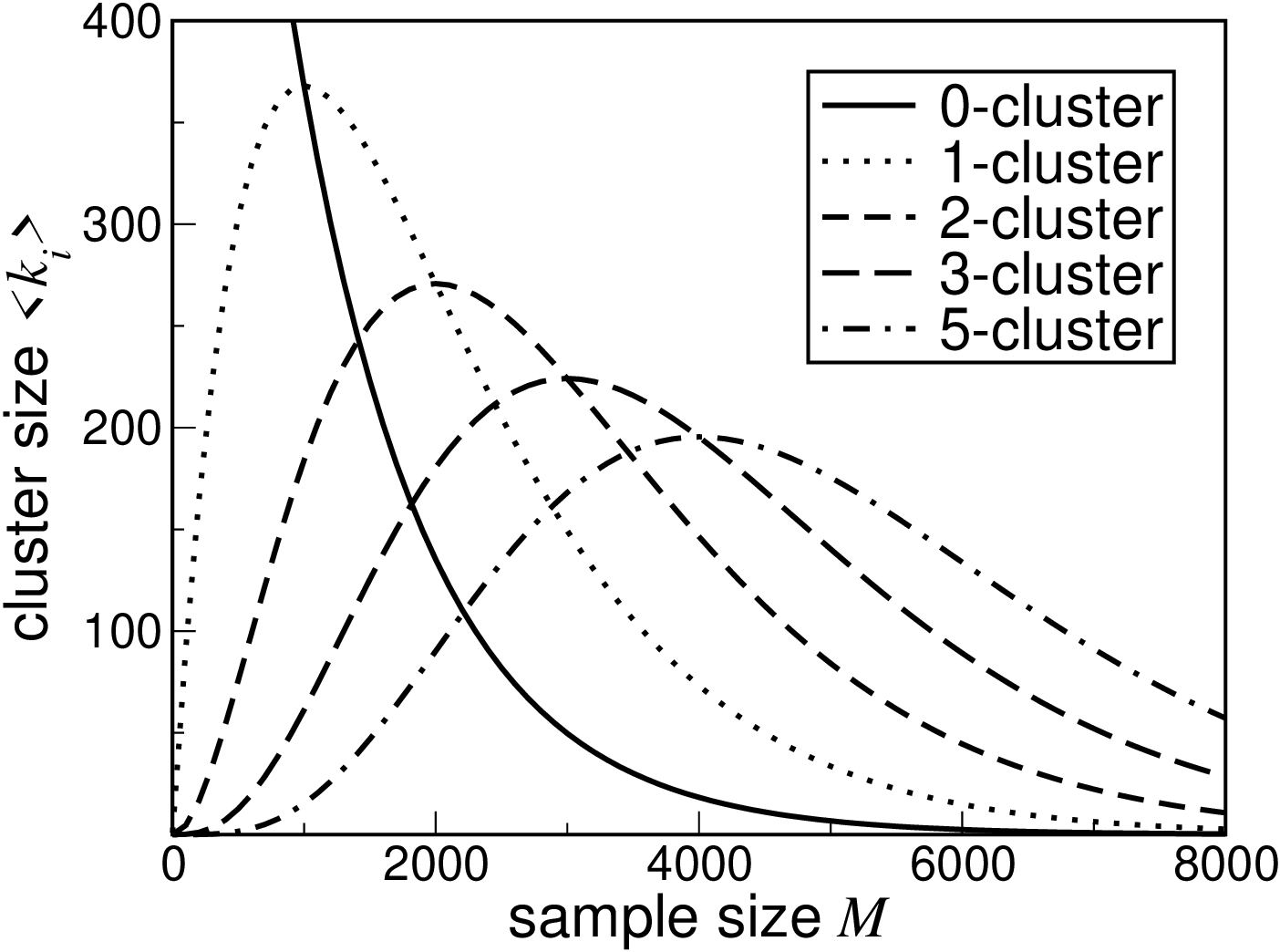}}
\caption{Expectation values $\left<k_i\right>$ for cluster sizes 
  $i=0,..,5$ over the sample size $M$ taken from a set of $N=1000$
  equidistributed species. The lines show the theoretical result Eq.
  \eqref{eq:Mom}. The symbols show the cluster sizes found by
  numerical simulations. The lower figure shows the same data for a
  larger range of the sample size $M$.}
\label{fig:Mom}
\end{figure}

Obviously, for small $M\ll N$, almost all of the $N$ species which
could be found in principle, belong to the 0-cluster, i.e., they do
not appear in a sample of size $M$. As $M$ increases the number of
single occupations $\left<k_1\right>$ increases as well, consequently
$\left<k_0\right>$ decays. For still growing $M$ the number of
multiple occupations becomes larger and, therefore, the sizes of the
0-cluster and 1-cluster decrease. Figure \ref{fig:Mom} shows the sizes
of the first clusters, $\left<k_0\right>$ to $\left<k_5\right>$, as a
function of the sample size $M$.  The lines show the theoretical
result Eq. \eqref{eq:Mom} and the symbols in the top of Fig.
\ref{fig:Mom} show the clusters as they have been found in numerical
simulations using equally distributed random numbers.

\begin{figure}
\centerline{\includegraphics[width=7cm,clip]{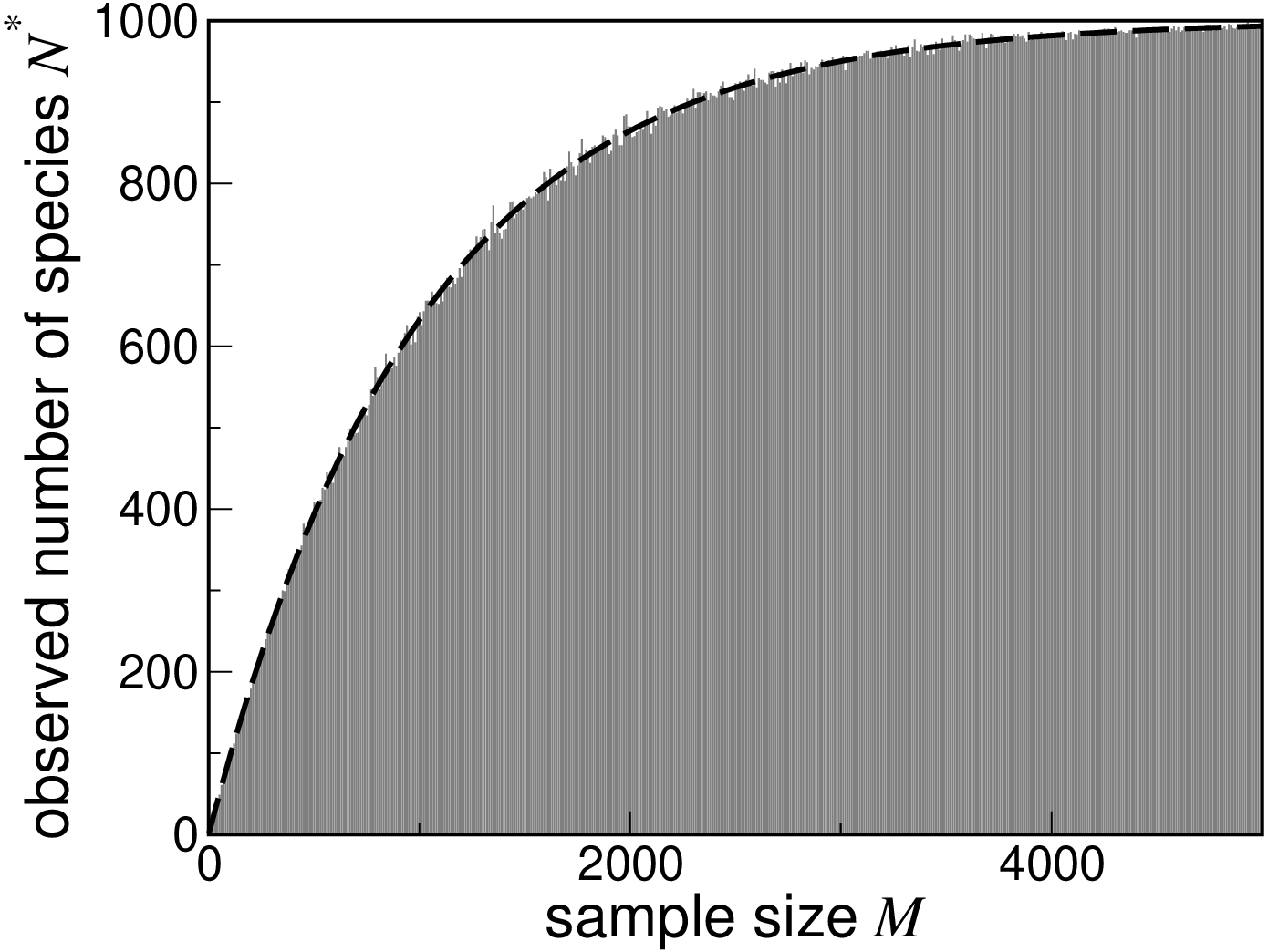}}
\caption{Number of species $N^{*}$ found in a sample of size $M$ 
  if $N=1000$ species occur all with the same probability
  $c_i=1/1000$. The dashed line shows the analytical result Eq.
  \eqref{eq:Nstar}, the impulses show the results of a computer
  simulation.}
\label{fig:Nstar}
\end{figure}

As a special case $\left<k_0\right>$ allows to determine the number of
different species $N^*$, which are expected to be found in a sample of
size $M$. This number is given by the total number of species $N$
minus the number of species which we expect to find with zero
representatives:
\begin{equation}
N^{*}=N - \left<k_0\right>\, ,
\end{equation}
i.e.,
\begin{equation}
\frac{N^{*}}{N}=1-\left(1-\frac{1}{N}\right)^M\,. 
\label{eq:Nstar}
\end{equation}
Figure \ref{fig:Nstar} shows the corresponding simulation results for
$N=1000$. For sample size $M=5000$ we notice that the average number
of found species is $N^{*}\approx 993$, i.e., on average about 7
species are not found. For $M=8000$ the average number of found
species is $N^{*}\approx 999.67$, here we can be optimistic to have
found at least one representative of all species.
For practical purposes it may be useful to note that even for rather
small values of $N$, Eq. \eqref{eq:Nstar} can be approximated with
very good accuracy in the entire range of $M$ by
\begin{equation}
\frac{N^{*{\rm  approx}}}{N}\approx 1-\exp \left(-\frac{M}{N}\right)\,. 
\label{eq:Nstarapprox}
\end{equation}
The maximal {\em absolute} deviation is $N^{*}-N^{* {\rm
    approx}}=1/e\approx 0.37$ which falls rapidly to $1/(2e)\approx
0.18$ as $N$ goes to infinity.

Using Eq. \eqref{eq:Mom} for the expectation values $\left<k_i\right>$
we obtain directly the observed rank-ordered distribution of relative
frequencies \cite{TPJanf,TPJanfEqui}:
\begin{equation}
  r^*_i=\left\{\!\!\begin{tabular}{lll}
 0 &\hspace{-0.1cm}for   &\hspace{-0.2cm}$N\ge i > N-\left<k_0\right>$\\ 
  $1/M$ &\hspace{-0.1cm}for   &\hspace{-0.2cm}$N-\left<k_0\right> \ge i  
> N-\left<k_0\right> -\left<k_1\right>$ \\
      &\hspace{-0.1cm}\dots&\\
  $i/M$ &\hspace{-0.1cm}for &\hspace{-0.2cm}$N-\sum\limits_{s=0}^{i-1}
      \left<k_s\right> \ge i > N-\sum\limits_{s=0}^{i} \left<k_s\right>$
    \end{tabular}
  \right. \label{eq:Hauf}
\end{equation}
Using Stirling's formula to expand the expressions in Eq.
\eqref{eq:Mom} the analytical result Eq. \eqref{eq:Hauf} can be
written easily in elementary functions.

\begin{figure}[htb]
\centerline{\includegraphics[width=7cm,clip]{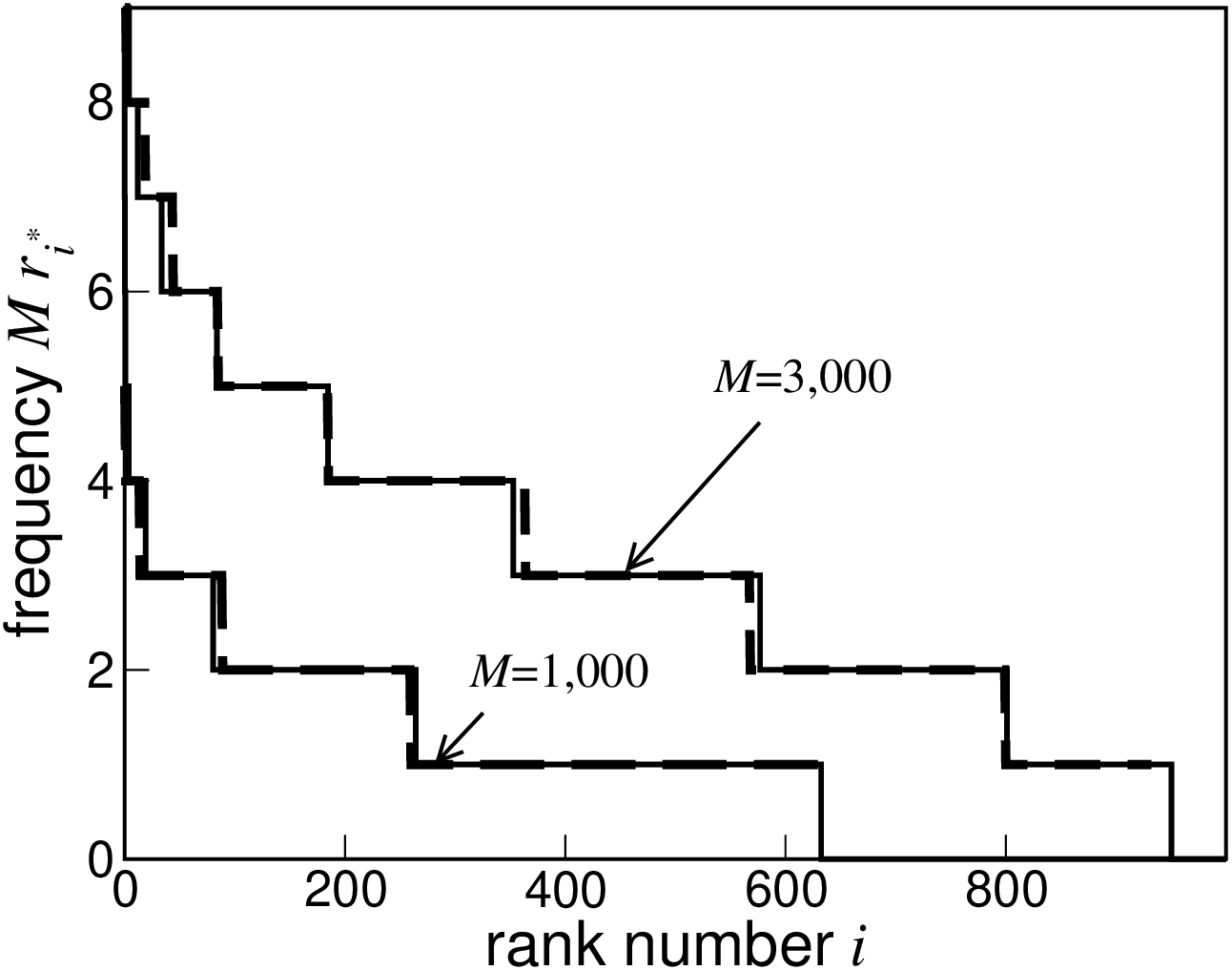}}
\centerline{\includegraphics[width=7cm,clip]{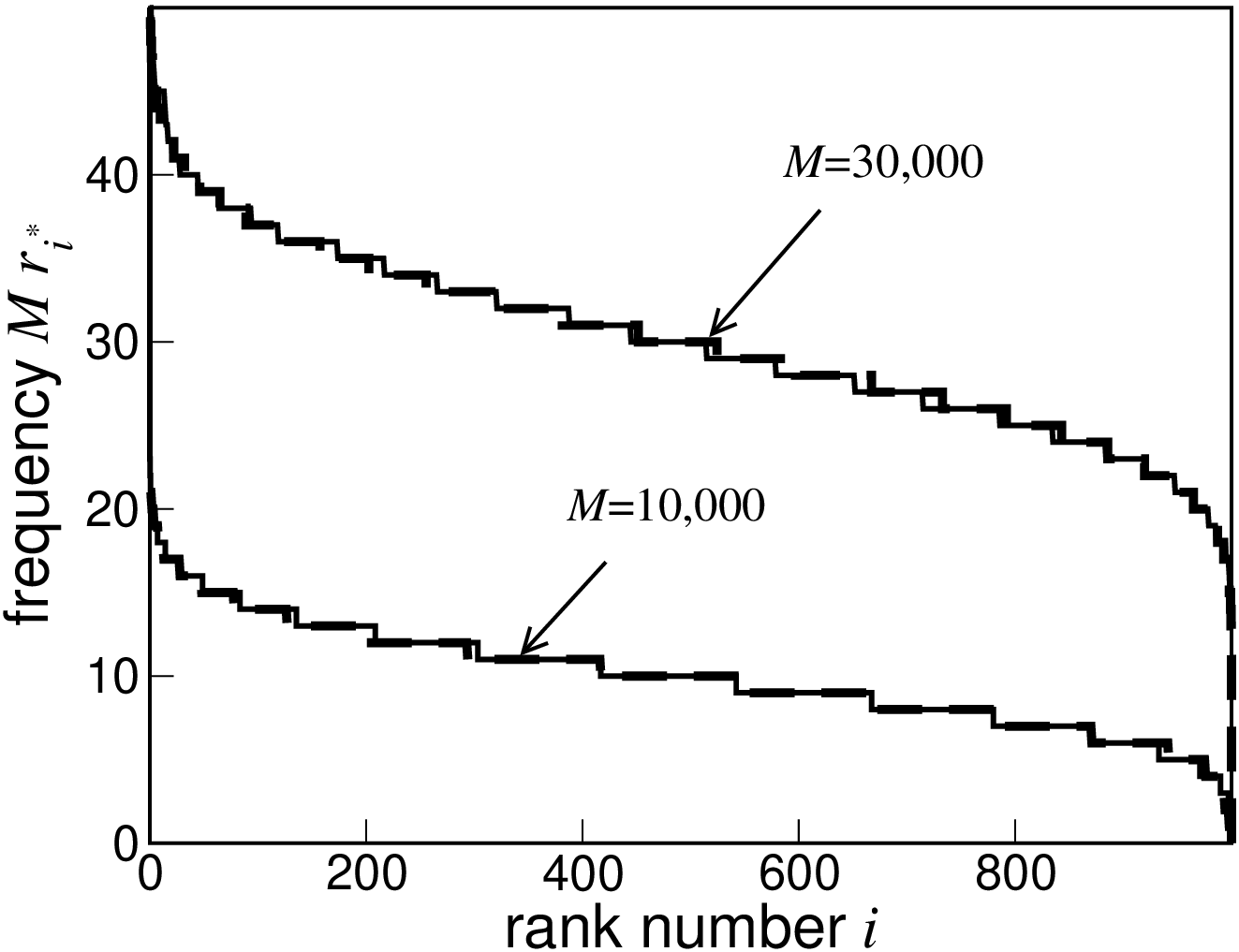}}
\caption{Numerically determined rank-ordered relative frequencies 
  $r_i^*$ for samples of size $M$ (dashed lines). The $N=1,000$
  species occur with equal probability, $p_i=1/1000$. The theoretical
  curves due to Eq. \eqref{eq:Hauf} are drawn with solid lines.}
\label{fig:Zipf}
\end{figure}

Figure \ref{fig:Zipf} shows rank-ordered relative frequencies
calculated from a sample of random numbers (dashed lines) together
with the theoretical distributions due to Eq.  \eqref{eq:Hauf} (solid
lines). (To plot more than one curve in the same figure we show the
absolute frequencies $Mr_i^*$, i.e., what is shown are the absolute
numbers of occurrence of species $i$ in a pool of size $M$.) The
combinatorial theory sketched above predicts the rank-ordered relative
frequencies which results from an equi-probability distribution with
good accuracy.

The figure demonstrates that indeed we are able to predict
analytically the rank-ordered observed frequencies which appear when a
sample of size $M$ is drawn from $N$ equally distributed species.  We
may turn the question around and answer the question: Which sample
size is required to make sure that that at least $N^*$ out of $N$
species are observed. The required sample size is in good
approximation
\begin{equation}
  \label{eq:zeta}
M = N \log \frac{N}{N-N^*} = - N \log \zeta\,.
\end{equation}
Here $\zeta$ is the percentage of species which is admittedly is not
to be represented in the sample. If we admit that about $5\%$ are not
represented we find, e.g., $M \simeq 2 N$, i.e. the size of the sample
should be about double the estimated species number.  This estimate
may be important for the planning of observations of nearly equally
probable species.

\subsection{General distributions}

\subsubsection{An alternative motivation of Eq. \eqref{eq:Mom}}
\label{sec:alternative} The derivation of Eq. \eqref{eq:exact} for the case of
equidistributed species requires some algebra and in this work we will
not derive a corresponding equation for general distributions. For the
equidistribution the order of the rank-ordered empirical frequencies
always corresponds the order of the true probabilities since all true
probabilities are identical, i.e., reordering the true probabilities
leads always to the equidistribution. This is different in the general
case: Due to fluctuations it may happen that $p_i^*<p_j^*$ although
$p_i>p_j$. The probability for the species to change ranks in the
empirical distribution depends on the difference $p_i-p_j$ (the larger
the difference the less probable they change ranks due to
fluctuations) and on the sample size (the larger the sample size the
less are the fluctuations, hence, the smaller the probability to
change ranks). A comprehensive calculation must take these exchange
probabilities into account.

Nevertheless, we wish to present an hypothesis which can be checked by
numerical simulations. We will demonstrate that although the
theoretical derivation of $\left<k_i\right>$ for the case of a
non-uniform probability distribution is somewhat simplified, the
predicted results agree well with numerics.

Let us discuss an alternative motivation of Eq. \eqref{eq:Mom}: Assume
there are $N$ species occurring with the same probability
$p_1=p_2=\dots=p_N=1/N$. The probability to find exactly $i$
representatives of species $j$ in a community of $M$ individuals, is
given by the binomial distribution
\begin{equation}
  \label{eq:binom}
  P_j(i)=\left(M\atop i\right) p_j^i\left(1-p_j\right)^{M-i}\,.
\end{equation}
The probability to find {\em any} species exactly $i$ times in a
community of size $M$ is the union of species 1 appearing $i$ times,
species 2 appearing $i$ times, etc. Since these events do not exclude
each other for $i<N/2$ one cannot sum directly the probabilities.
Instead, one has to apply the inclusion-exclusion principle \cite{JK}
to subtract the intersection probabilities which in fact has been done
to derive Eq. \eqref{eq:Mom}, see \cite{JT,TJ}. Let us see what
happens if we ignore the intersection probabilities: The expectation
value for the number of species which appear in a sample of size $M$
exactly with $i$ representatives reads then
\begin{eqnarray}
  \label{eq:binomExp}
  \left<k_i\right>&=&\sum\limits_{j=1}^N P_j(i) = N \left(M\atop i\right) 
\left(\frac1N\right)^i\left(1-\frac1N\right)^{M-i}\nonumber\\
&=& \left(M\atop i\right) N^{(1-i)}\left(1-\frac1N\right)^{M-i}\!\!\!\!\!\,,
\end{eqnarray}
which is identical with Eq. \eqref{eq:Mom}. We want to point out again
that the derivation of the first moments is incomplete but it yields
the correct result. In contrast to the exact derivation this simple
motivation for the equidistribution has the great advantage that it
can be generalized to the case of an arbitrary distribution. In
general, according to Eqs. \eqref{eq:binom} and \eqref{eq:binomExp},
the expectation value for the number of species which appear in a
sample of size $M$ exactly $i$ times is
\begin{equation}
  \label{eq:binomExpUnfair}
  \left<k_i\right>= \sum\limits_{j=1}^N\left(M\atop i\right) p_j^i\left(1-p_j\right)^{M-i}\,.  
\end{equation}
It can be easily checked that this distribution has the correct
normalization imposed by Eqs. \eqref{eq:condition1} and
\eqref{eq:condition2}.

Having always in mind that we have no rigorous proof for the
correctness of this result yet, we want to check its validity by
numerical simulations.

\subsubsection{Example: step-wise equidistribution}

\begin{figure}[htbp]
  \centerline{\includegraphics[width=7cm,clip]{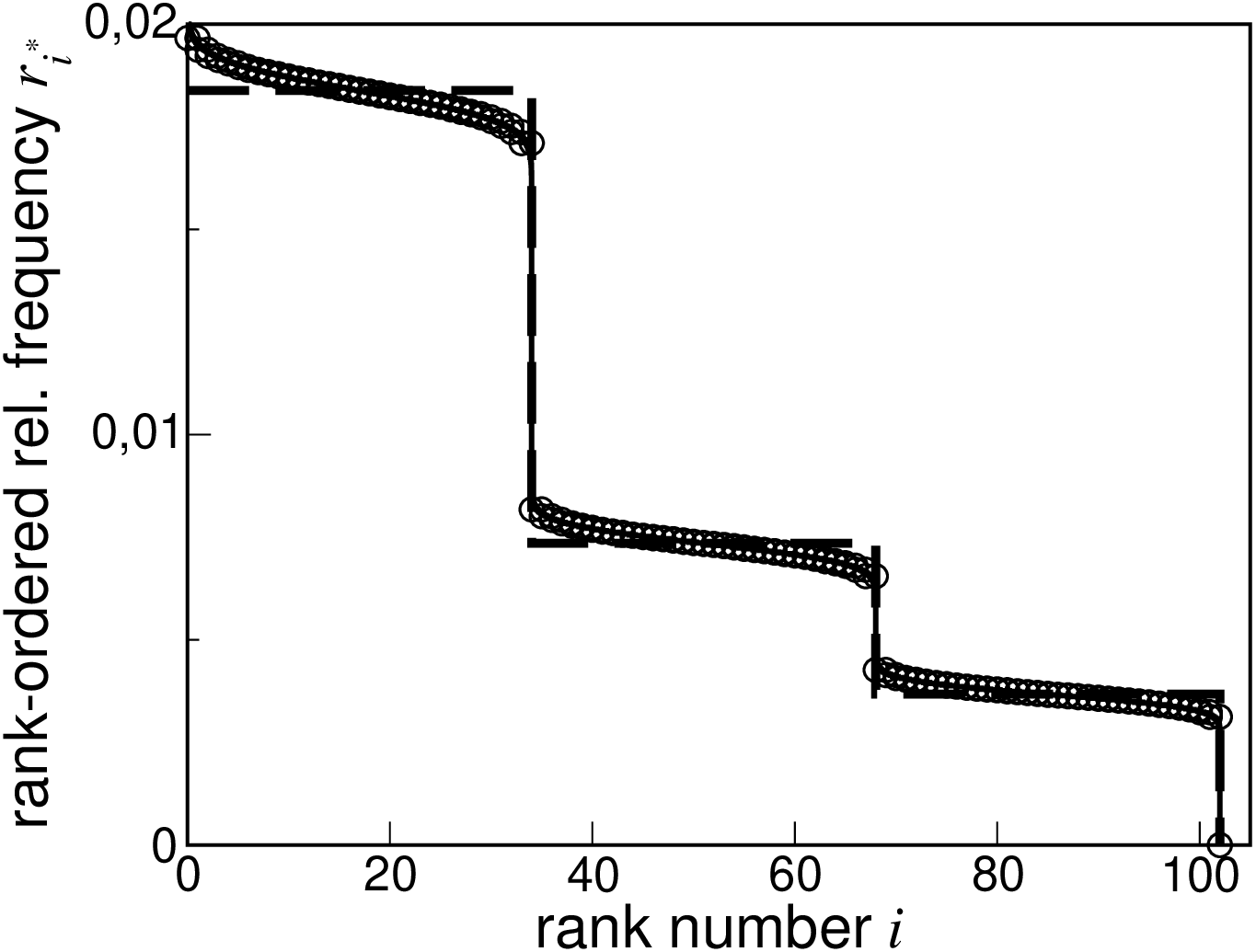}}
  \centerline{\includegraphics[width=7cm,clip]{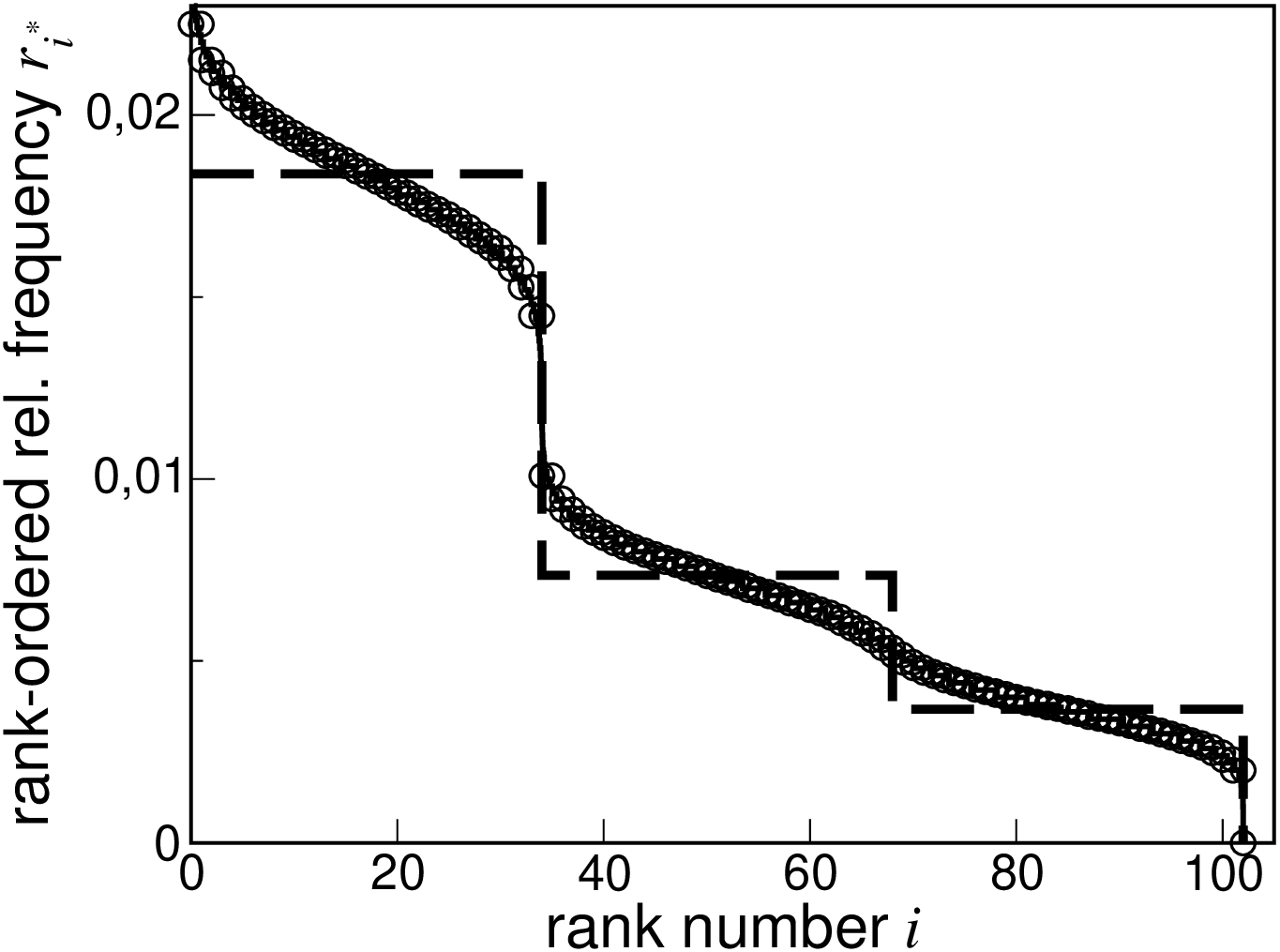}}
  \centerline{\includegraphics[width=7cm,clip]{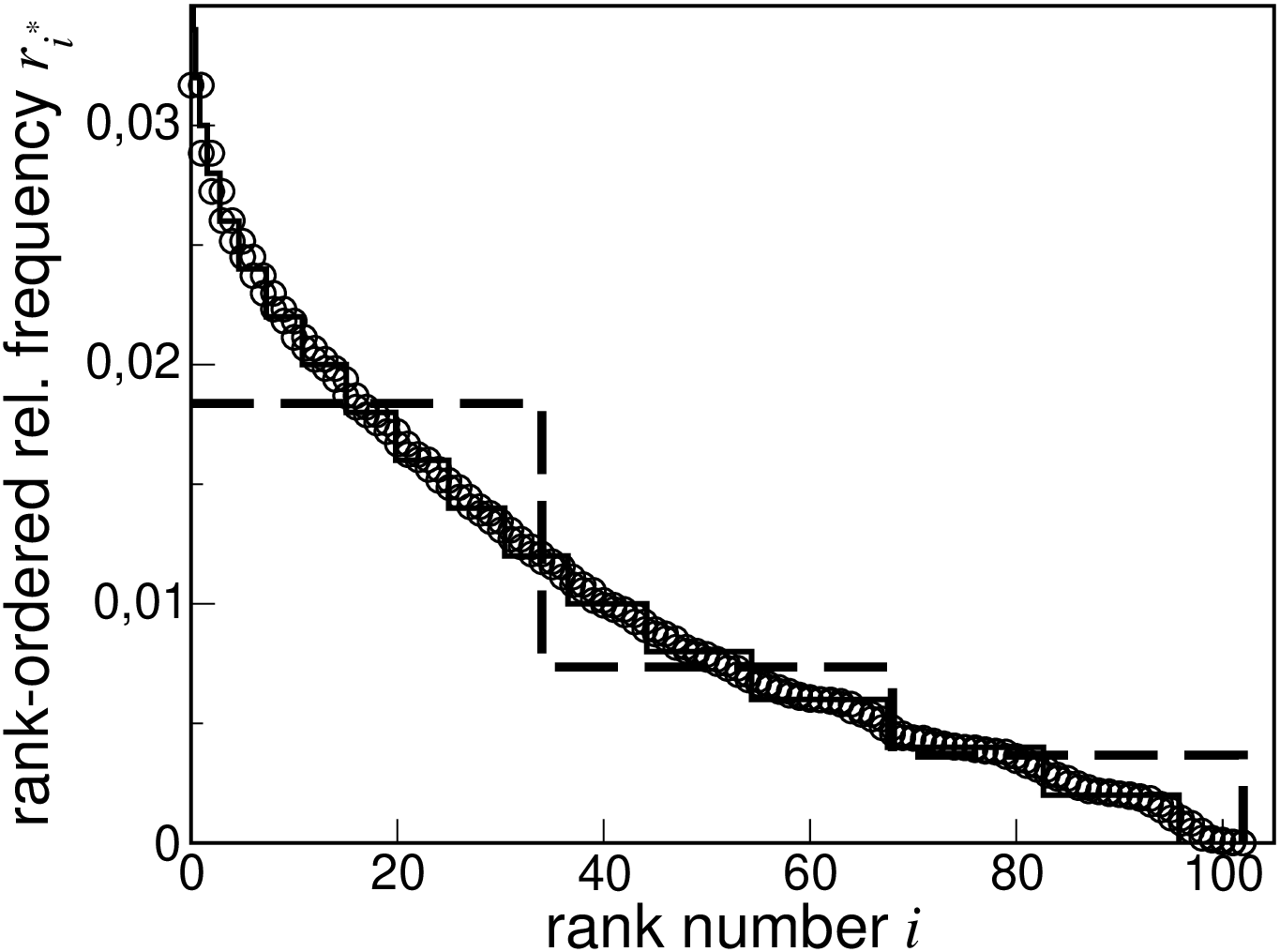}}
  \centerline{\includegraphics[width=7cm,clip]{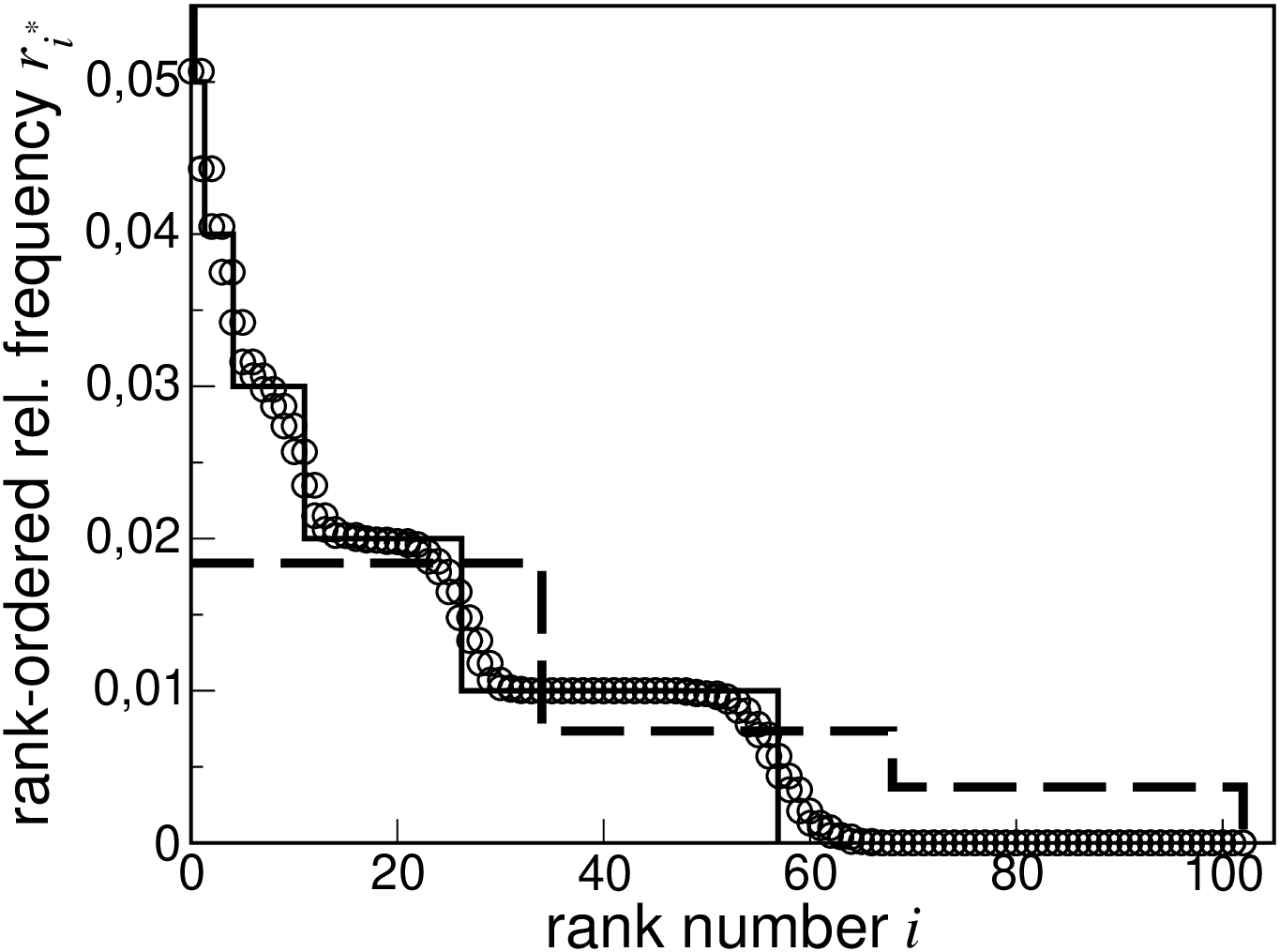}}
  \caption{Rank-ordered relative frequencies of step-wise equidistributed 
    species due to Eq. \eqref{eq:unfairdie} for sample sizes
    $M=50,000$, $M=5,000$, $M=500$, and $M=100$ (top to bottom). The
    dashed lines show the probabilities due to Eq.
    \eqref{eq:unfairdie}, the full lines show the predicted relative
    frtequencies due to Eq. \eqref{eq:conc} and the circles show the
    results of a Monte Carlo simulation.}
  \label{fig:steps}
\end{figure}

We wish to demonstrate the application of Eq.
\eqref{eq:binomExpUnfair} using a step-wise equidistribution of
$N=102$ species.  Let us assume for the probabilities:
\begin{equation}
  \label{eq:unfairdie}
  r_i=p_i=\left\{
    \begin{tabular}{llrl}
$p_\alpha=15/(8N)$ & for & $1\!\!\!\!\!$    &$\le i\le N/3$\\ 
$p_\beta=6/(8N)$   & for & $N/3\!\!\!\!\!$  &$< i\le 2N/3$\\ 
$p_\gamma=3/(8N)$  & for & $2N/3\!\!\!\!\!$ &$< i\le N$\,.
    \end{tabular}\right.
\end{equation}
The normalization can be checked easily: $\sum_{i=1}^N p_i=1$. For
these probabilities of the species we obtain from Eq.
\eqref{eq:binomExpUnfair}
\begin{eqnarray}
  \label{eq:ejemplo}
\left<k_i\right> &=& \left(M \atop i\right) \left[ \frac N3 p_\alpha^i
(1-p_\alpha)^{M-i}+\right.\\ &+&\left.\frac N3 p_\beta^i
(1-p_\beta)^{M-i}+\frac N3 p_\gamma^i (1-p_\gamma)^{M-i} \right]\nonumber\,.
\end{eqnarray}
The expected rank-ordered empirical relative frequencies, $r^*_i$, can
be found from Eq. \eqref{eq:Hauf} in the same way as previously:
$\left<k_0\right>$ is the number of species which on average will not
be found in a sample of size $M$, $\left<k_1\right>$ is the number of
species which will appear with one representative, $\left<k_2\right>$
species are with two representatives each, etc. and finally
$\left<k_M\right>$ is the number of species which are expected to be
found with $M$ representatives. Obviously, no species can appear with
more than $M$ representatives since our sample is of size $M$. To
generate the rank-ordered observed relative frequencies we notice that
on average these values jump from $(i+1)/N$ to $i/N$ at rank-positions
$N-\sum\limits_{s=0}^{i-1} \left<k_i\right>$. Hence, the expected
empirical relative frequencies are
\begin{equation}
  \label{eq:conc}
r^*_i\!=\!\left\{
  \begin{tabular}{llrl}
    \!\!\!0         & \!\!for\!\! & $N$           & $\!\!\!\!\!\ge i >N-\left<k_0\right>$\\
    \!\!\!$1/M\!\!$ & \!\!for\!\! & $N-\left<k_0\right>$       & $\!\!\!\!\!\ge i >N-\left<k_0\right>-\left<k_1\right>$\\
    \!\!\!$2/M\!\!$ & \!\!for\!\! & $\!\!\!\!\!N-\left<k_0\right>-\left<k_1\right>$   & $\!\!\!\!\!\ge i>$\\
&&&$\!\!\!\!\!\!\!\!\!\!\!\!\! >N-\left<k_0\right>-\left<k_1\right>-\left<k_2\right>$\\ 
\dots\\ 
\!\!\!    $1$       & \!\!for\!\! & $N-\sum_{s=0}^{M} \left<k_s\right>$ & $\!\!\!\!\!\ge i > 0$
  \end{tabular}
\right.
\end{equation}
Note that, in general, the average cluster sizes $\left<k_s\right>$
and, therefore, $i$ are not integers. To check this formula, in Fig.
\ref{fig:steps} we show the true probability distribution due to Eq.
\eqref{eq:unfairdie} (dashed lines), the prediction of the observed
relative frequencies due to Eq. \eqref{eq:conc} (solid lines) and the
results of a Monte Carlo simulation (circles), where the data have
been averaged over 100 independent drawings of random numbers. The
analytical and numerical results agree with good accuracy.

\subsubsection{Example: exponential distribution}
\begin{figure}[htbp]
\centerline{\includegraphics[width=7cm,clip]{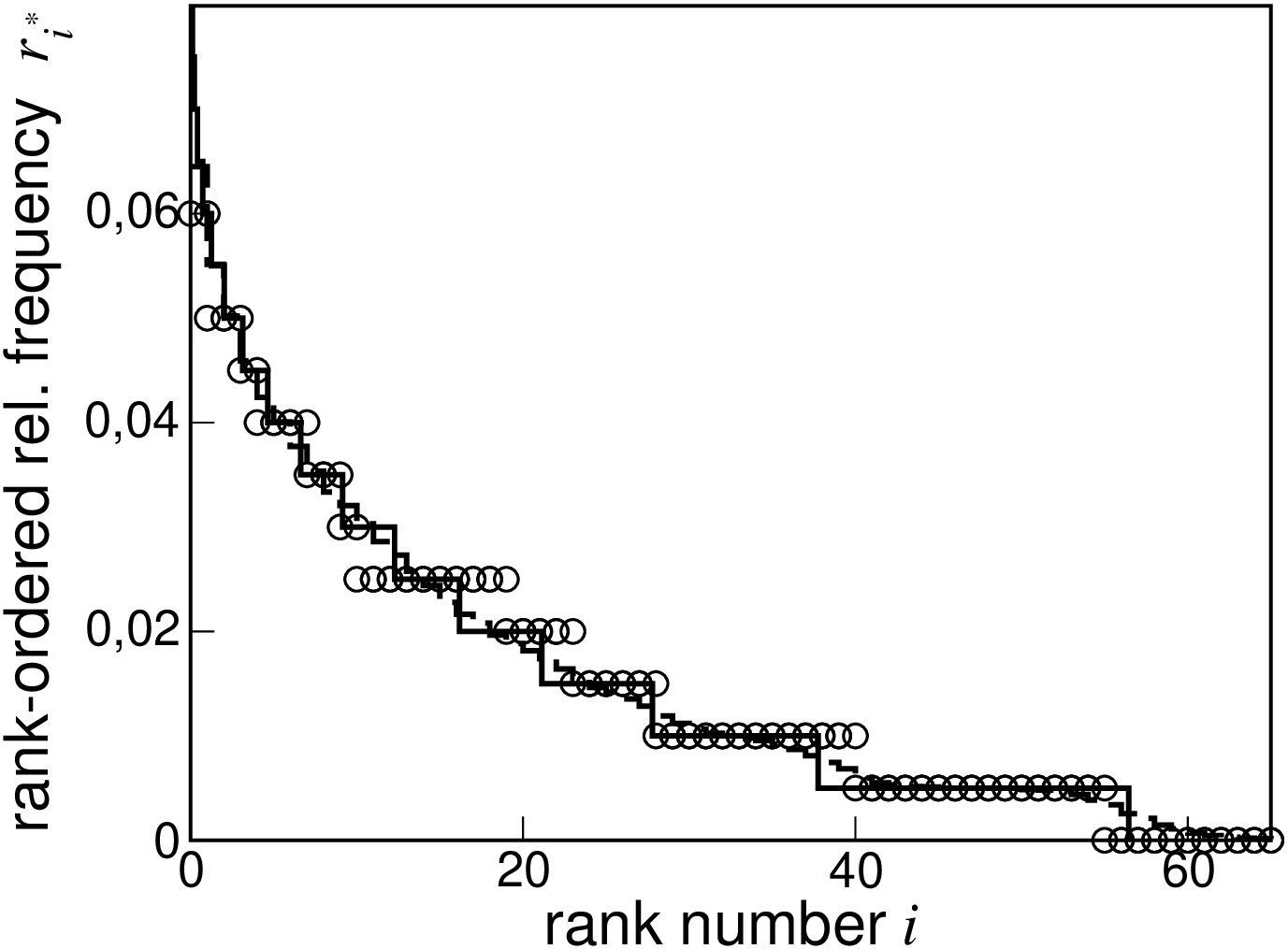}}
\centerline{\includegraphics[width=7cm,clip]{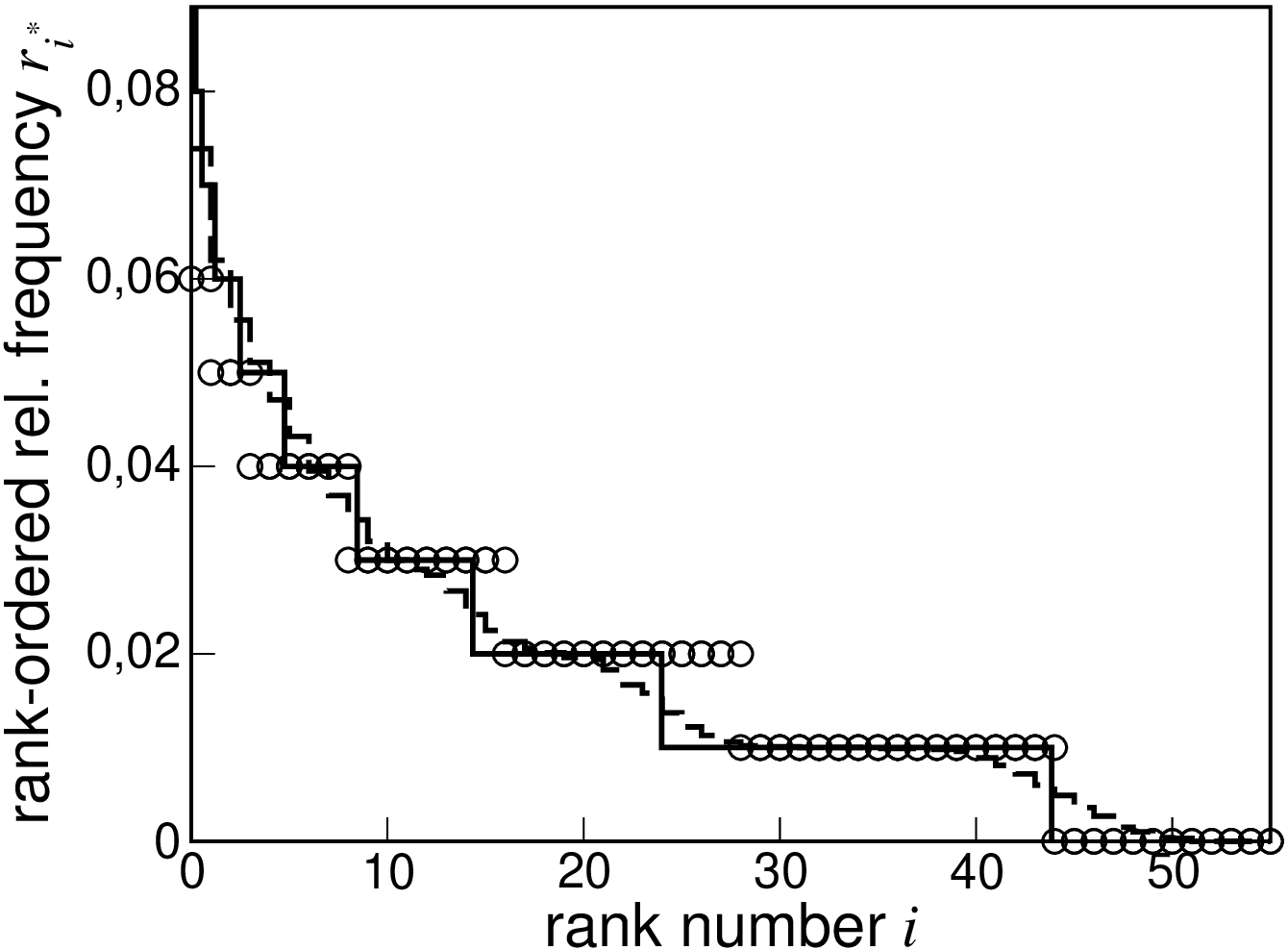}}
\centerline{\includegraphics[width=7cm,clip]{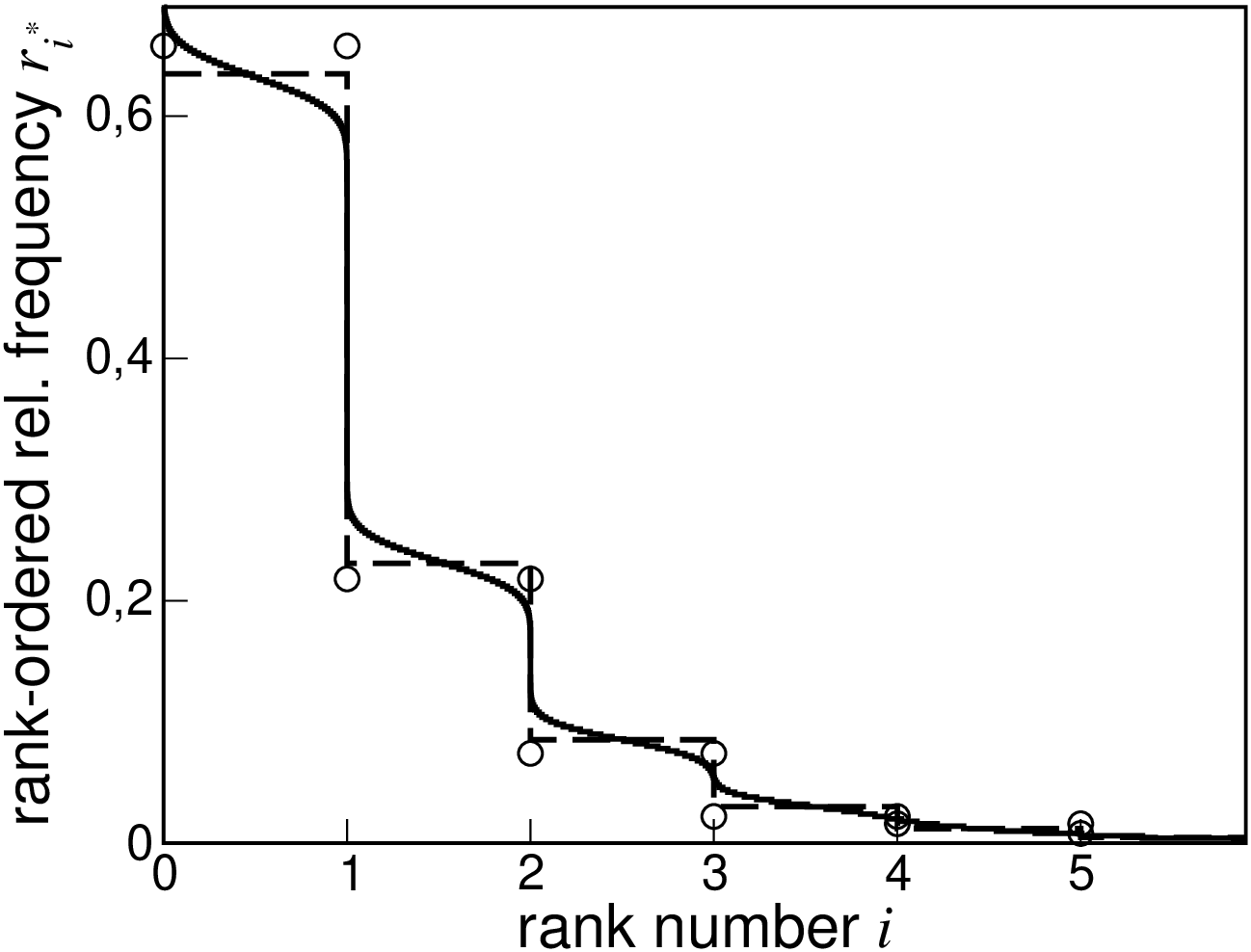}}
\caption{Rank-ordered relative frequencies for exponentially distributed 
  species as defined by Eq. \eqref{eq:defexp} with $\alpha=0.05$,
  $N=100$ and $M=200$ (top), $\alpha=0.05$, $N=200$, $M=100$ (middle),
  $\alpha=1.0$, $N=100$, $M=500$ (bottom). The theoretical results
  (solid lines) are due to Eqs. \eqref{eq:conc} and \eqref{eq:Kexp},
  the circles show the distribution of a single set of $M$ random
  numbers from the interval $[1\dots N]$ drawn due to the distribution
  Eq. \eqref{eq:defexp}, and the dashed lines show averages over 100
  such experiments.}
  \label{fig:05100200}
\end{figure}

As a second example we wish to check the validity of Eq.
\eqref{eq:binomExpUnfair} by means of a (shifted) exponential
probability distribution
\begin{equation}
r_i=p_i=\frac{\alpha}{1-\exp(\alpha N)} \exp(-\alpha i)
\label{eq:defexp}
\end{equation}
with $0\le i\le N$, i.e., $\int_{i=0}^N p_i = 1$. From Eq.
\eqref{eq:binomExpUnfair} we obtain
\begin{eqnarray}
\left<k_i\right>= && \frac{M!}{i! (M-i)!} \left( \frac{1-r_i}{1-r_i^{N}}\right )^i \times\nonumber \\ 
&& \times \sum_{j=1}^N  \, r_i^{i(j-1)} \left(1-\frac{1-r_i}{1-r_i^{N}} r_i^{j-1} \right)^{M-i}  \,.
\label{eq:Kexp}
\end{eqnarray}
Figure \ref{fig:05100200} shows the theoretical predictions together
with results of numerical simulations. Again, theory agrees well with
the numerical results.
Due to the excellent agreement of the Eq. \eqref{eq:Kexp} with
numerics we conclude that Eq. \eqref{eq:binomExpUnfair} allows to
predict the observed relative frequencies, provided the true
probability distribution is given.

\section{Inferring probabilities from experiments}
\subsection{How to determine probabilities?}
\label{sec:How}

In strict sense, probabilities cannot be determined by experiments for
an obvious reason: even for a fair die it is mathematically possible,
although not very probable, to cast the die 1000 times and to find
1000 times the six. From such a measurement, of course, one would
hardly conclude that the die is fair, i.e., that the sides one to six
appear with equal probability. Hence, we have to require that the
measurement is {\em representative}. The strict definition of this
term is not easy since for $M=10$ both measured sequences,
$5-5-3-4-2-6-5-6-1-3$ and $6-6-6-6-6-6-6-6-6-6$ occur with equal
probability. The great advantage of working with rank-ordered
measurements is that the order of the measured sequence is irrelevant,
i.e., the measurement $5-5-3-4-2-6-5-6-1-3$ would lead to the same
measured rank-ordered frequencies as $2-6-6-3-4-5-5-3-1$ or
$5-5-5-3-3-6-6-1-2-4$. In this sense, a rank-ordered measurement from
a sequence $5-5-3-4-2-6-5-6-1-3$ represents many more possible
measured configurations than $6-6-6-6-6-6-6-6-6-6$. Similar as in
statistical mechanics for the derivation of the canonical distribution
we will call a measurement {\em representative} if there are many
equivalent measurements (permutations) which all belong to the same
rank-ordered sequence.

In strict mathematical sense we have to repeat the experiment of
drawing a sample of size $M$ an infinite number of times in order to
get an averaged and representative set of rank-ordered frequencies. If
we, however, had all these measurements the method presented in this
article would turn out to be meaningless since for an infinite set of
measurements the observed probabilities approximate the true ones, see
Eq. \eqref{cdef}. Following the same argumentation, in order to
measure the pressure of air in a room we would also need an infinite
set of measurements since there is a non-zero probability (although
never observed under common conditions) that all air molecules are
located in one half of the room and our manometer would show the
double pressure or zero, depending on which half of the room is
populated. Therefore, there is not much difference between measuring
the pressure of air and inferring probabilities from measurements: in
both cases one relies on the fact that a {\em representative}
measurement is, by definition, a very probable one.

\subsection{Optimization of cluster distributions}
\label{firstmethod}

Equation \eqref{eq:binomExpUnfair} allows to predict in a systematic
way the expectation values of the clusters sizes $\left<k_i\right>$,
$i=0\dots M$, provided the probabilities $p_i$, $i=1\dots N$, are
known.  Note that the average cluster sizes $\left<k_i\right>$ based
on the species-ordered probabilities $p_i$ are identical with those
based on the rank-ordered probabilities $r_i$. We can say then that
Eq. \eqref{eq:binomExpUnfair} states the relation between the observed
rank-ordered frequencies $r^*_i$ and the rank-ordered probabilities
$r_i$. This relation permits to infer the rank-ordered probabilities
from a set of observed frequencies $r^*_i$.

In this section we propose a variational method to estimate the
distribution $\{r_i, i=1,\dots,N\}$ from data of a measurement.

Consider a set of experimentally determined cluster sizes $k_i^{\rm
  exp}$, $i=1\dots M$. This set can be determined by counting, how
many species in a sample of size $M$ appeared with one individual in
the sample ($k_1^{\rm exp}$), with two individuals ($k_2^{\rm exp}$),
etc.  We assume further that the set of experimentally determined
relative frequencies (and, therefore, cluster sizes) is {\em
  representative} in the sense as discussed in Section \ref{sec:How},
i.e., we assume that there exists a (unknown) probability distribution
which leads to the {\em averaged} cluster sizes
$\left<k_1\right>\approx k_1^{\rm exp}$, $\left<k_2\right>\approx
k_2^{\rm exp}$, etc.

Equation \eqref{eq:binomExpUnfair} establishing the relation between
the probabilities $r_i$ and the averaged cluster sizes
$\left<k_i\right>$ allows to construct a variational scheme. This is
done by constructing the dimensionless objective function
\begin{equation}
  \label{eq:minimize}
  \psi_{(k)}\equiv\sum\limits_{i=0}^M\left(\left<k_i\right>-k_i^{\rm exp}\right)^2.
\end{equation}
and requiring that it is minimal for the (unknown) set of
probabilities $r_i$. The index $(k)$ of $\psi_{(k)}$ indicates that
the objective function refers to the cluster distribution.

Starting from a trial initial set of rank-ordered probabilities $r_i$,
e.g. the equidistribution, and an initial trial number of species $N$,
e.g.  the observed number of species $N^*$ (implying that $k_0^{\rm
  exp}=0$), the probabilities can be approximated numerically by a
gradient method
\begin{equation}
  \label{eq:gradient}
  r_i:=r_i-\epsilon \frac{\partial \psi_{(k)}}{\partial r_i}\,,~~~~~~i=1,\dots,N\,,
\end{equation}
with $\epsilon$ being a small number. Using Eqs.
\eqref{eq:binomExpUnfair} and \eqref{eq:binom} we obtain
\begin{eqnarray}
  \frac{\partial \psi_{(k)}}{\partial r_i}
  &=& \sum\limits_{j=0}^M \frac{\partial \psi_{(k)}}{\partial 
\left<k_j\right>}\frac{\partial \left<k_j\right>}{\partial r_i}\\
&=& 2\sum\limits_{j=0}^M\left[ \left(\left<k_j\right>-k_j^{\rm exp}\right) 
\left(\frac{j}{r_i}-\frac{M-j}{1-r_i}\right) P_i(j)\right]. \nonumber
  \label{eq:gradient1}
\end{eqnarray}
The so modified rank-ordered probabilities $r_i$ have to be normalized 
\begin{equation}
\sum\limits_{i=1}^N r_i=\sum\limits_{i=1}^N p_i=1\,.  
\end{equation}
Equation \eqref{eq:gradient} and subsequent normalization has to be
applied till convergence of $\psi_{(k)}(N)$ is achieved. Of course,
the initial value $N$ might not be the true number of different
species, i.e., on top of the $r_i$ for fixed $N$ we have to optimize
the value of $N$ itself. This can be done by performing a sequence of
minimizations for different values of $N$ ranging from the observed
value $N^*$ till some $N_{\rm max}$. The result of this set of
minimizations is the function $\psi_{(k)}(N)$, which has to be minimum
for the most probable value of $N$.

For several examples we have been able to determine the probabilities
$r_i$ up to good accuracy. This method has, however, a drawback: for
the case of rather large sample size $M$, when the observed relative
frequencies approximate the probabilities, we expect that it is
simpler to infer probabilities from observed frequencies. Instead, for
increasing $M$ it becomes more and more difficult since the cluster
sizes $\left<k_i\right>$ become small. This can be seen, e.g., from
Fig. \ref{fig:zipf}: in the upper figure for $M=10^3$ the typical size
of the clusters is $k_i\sim 100$, whereas in the lower figure drawn
for $M=10^6$ the typical size is $k_i\sim 1$. Therefore, the larger
the sample size the larger become the fluctuations of the measured
cluster sizes $k_i^{\rm exp}$ and the expression in Eq.
\eqref{eq:gradient} becomes ill defined. Considering that a typical
cluster size is given by the number of observed different species
$N^*$ divided by the sample size $M$, the described method is useful
when $N^*/M \ge 1$. In this case it yields reliable results.

\subsection{Direct optimization of the probabilities}

To overcome the mentioned problem, the second proposed algorithm deals
directly with the rank-ordered distribution $r_i$ instead of the
cluster sizes $k_i$.  This method is very similar to a Monte Carlo
simulation in which the function to minimize is the deviation between
the predicted rank-ordered frequencies and the experimentally observed
frequencies.

Given an experimentally determined set of frequencies $M_i^{\rm exp}$
($i=1..N^{\rm exp}$), e.g., $M_1^{\rm exp}=25$, $M_2^{\rm exp}=15$,
$M_3^{\rm exp}=110$, etc., with $N^{\rm exp}$ being the observed
number of different species in the sample, the following algorithm
determines approximately the probabilities:

\begin{enumerate}
\item Determine the total number of individuals in the sample
  $M=\sum_{i=1}^{N^{\rm exp}} M_i^{\rm exp}$.
\item Order the frequencies according to their rank, i.e., $r_1^{\rm
    exp}=110$, $r_2^{\rm exp}=25$, $r_3^{\rm exp}=15$, etc.
\item Determine a trial initial value of the total number of species
  $N$, for example by means of Eq. \eqref{eq:Nstar}, i.e., determine
  the initial value of $N$ with the assumption that the (unknown)
  probabilities are identical.
\item Initialize the trial rank-ordered probabilities which are to be
  determined, for example, with $r_i=1/N$.
\item \label{label:loop} Predict the rank-ordered observed relative
  frequencies $r^*_i$, $i=1,\dots,N$ which are expected to be found
  when drawing a sample of $M$ individuals according to the trial
  probabilities. This can be done by two different methods, either by
  \begin{itemize}
  \item[(i)] calculating the expected cluster distribution due to Eq.
    \eqref{eq:binomExpUnfair} and then the rank-ordered frequencies
    via Eq. \eqref{eq:conc},
\item[(ii)] or by the following procedure
    \begin{enumerate}
    \item[(a)] draw $M$ random numbers from the interval $[1,N]$ with
      probabilities $r_i$ using a Metropolis algorithm,
  \item[(b)] count the occurrences of the numbers $1,2,\dots,N$ and
    sort these frequencies due to their rank,
  \item[(c)] repeat steps (a) and (b) a number of times, e.g. 10, and
    average the rank-ordered distributions. (Note that it is
    essential, first to rank-order and then to average.)
    \end{enumerate}
  \end{itemize}
\item Determine the deviation of the experimentally observed
  rank-ordered frequencies $\{r_i^{\rm exp}\}$ and the predicted
  rank-ordered frequencies $\{r_i^*\}$
  \begin{equation}
    \label{eq:deviat}
    \psi_{(r)} = \sum\limits_{i=1}^N \left|r_i^{\rm exp} - r_i^*\right|\,.
  \end{equation}
  The index $(r)$ indicates that $\psi_{(r)}$ is computed based on the
  frequencies $r_i$.\footnote{Formally Eq. \eqref{eq:deviat} is not
    perfectly correct. Since the indices $i$ in the distribution
    $\{r_i^*\}$ are no integers (see above), $\chi_{(r)}$ has to be
    computed as the integral difference between two functions with $i$
    being the integration variable. Since the precise mathematical
    notation appears to be cumbersome without contributing to deeper
    understanding we leave Eq. \eqref{eq:deviat} in its present form.}
\item \label{label:min} Modify the probabilities $r_i$ in order to
  minimize $\psi_{(r)}$.
\item proceed with item \ref{label:loop} until the deviation
  $\psi_{(r)}$ is sufficiently small or until no further progress can
  be achieved.
\end{enumerate}

\noindent The critical step is item \ref{label:min} when the probabilities are
modified. This has been done either in a deterministic way similar to
Eq.  \eqref{eq:gradient}, or by proposing a Monte Carlo trial movement
in the rank-ordered frequencies $r_i\to r_i+\Delta r_i$, with $\Delta
r_i$ being some random number and subsequent normalization. This
change is accepted if $\psi_{(r)}\left(r_i+\Delta r_i\right)\le
\psi_{(r)}\left(r_i\right)$, otherwise it is rejected. Both methods
$(i)$ and $(ii)$ yield very similar results. In this step the value of
the total number of species $N$ has to be also modified: after $t$
trial movements of the frequencies we propose a trial number of
species $N\to N+\Delta N$, with $\Delta N$ being some random number
such that $N$ keeps smaller than the initial $N$ corresponding to a
uniform distribution. This movement is accepted if $\psi_{r)}(N+\Delta
N)\le \psi_{(r)}(N)$, otherwise it is rejected.  Alternatively the
optimization could be performed for several values of $N$, as proposed
in Sec. \ref{firstmethod}, in order to find the minimum of the
function $\psi_{(r)}(N)$.

\medskip

\begin{figure}[htbp]
  \centerline{\includegraphics[width=7cm,clip=]{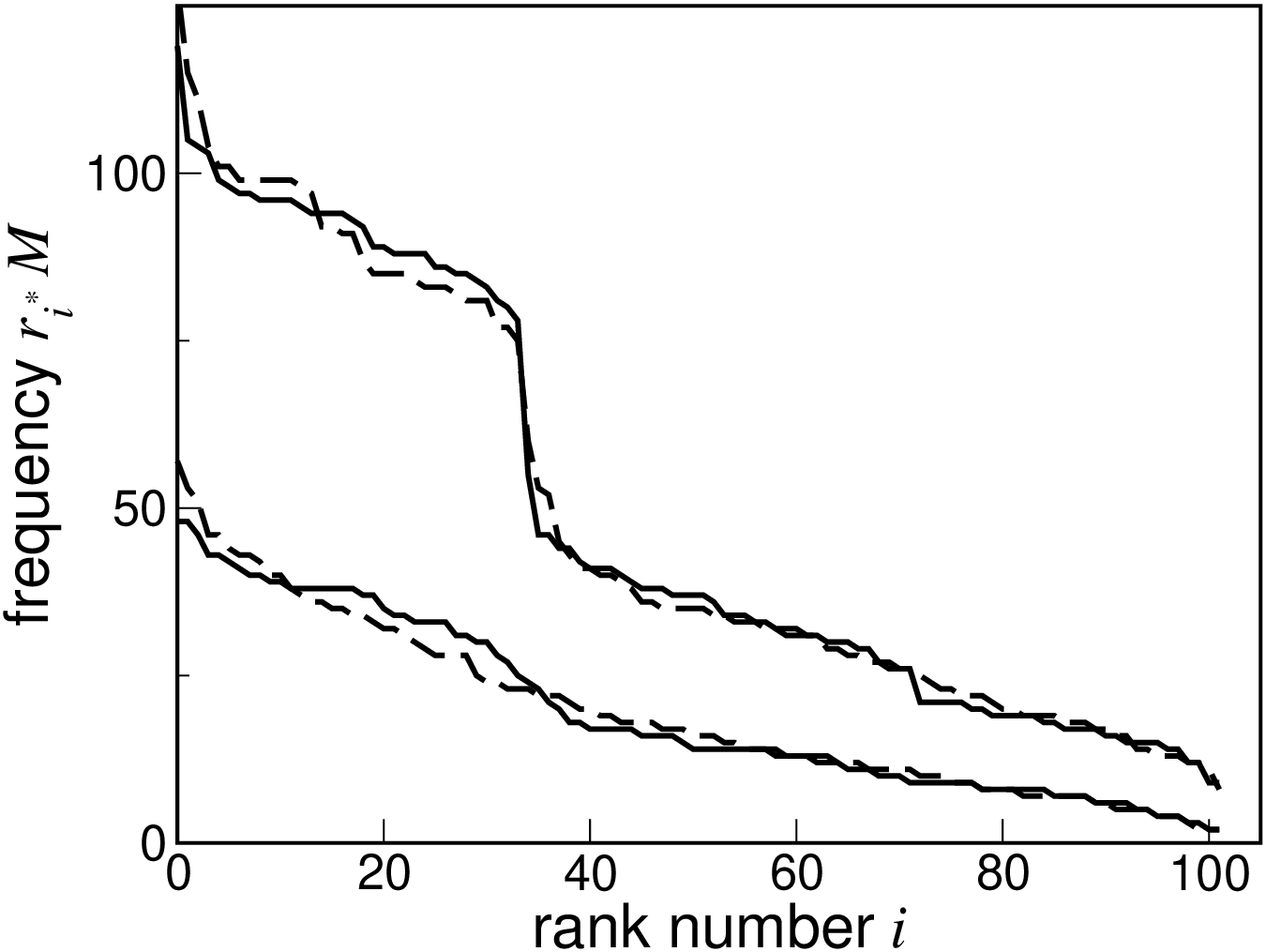}}
  \centerline{\includegraphics[width=7cm,clip=]{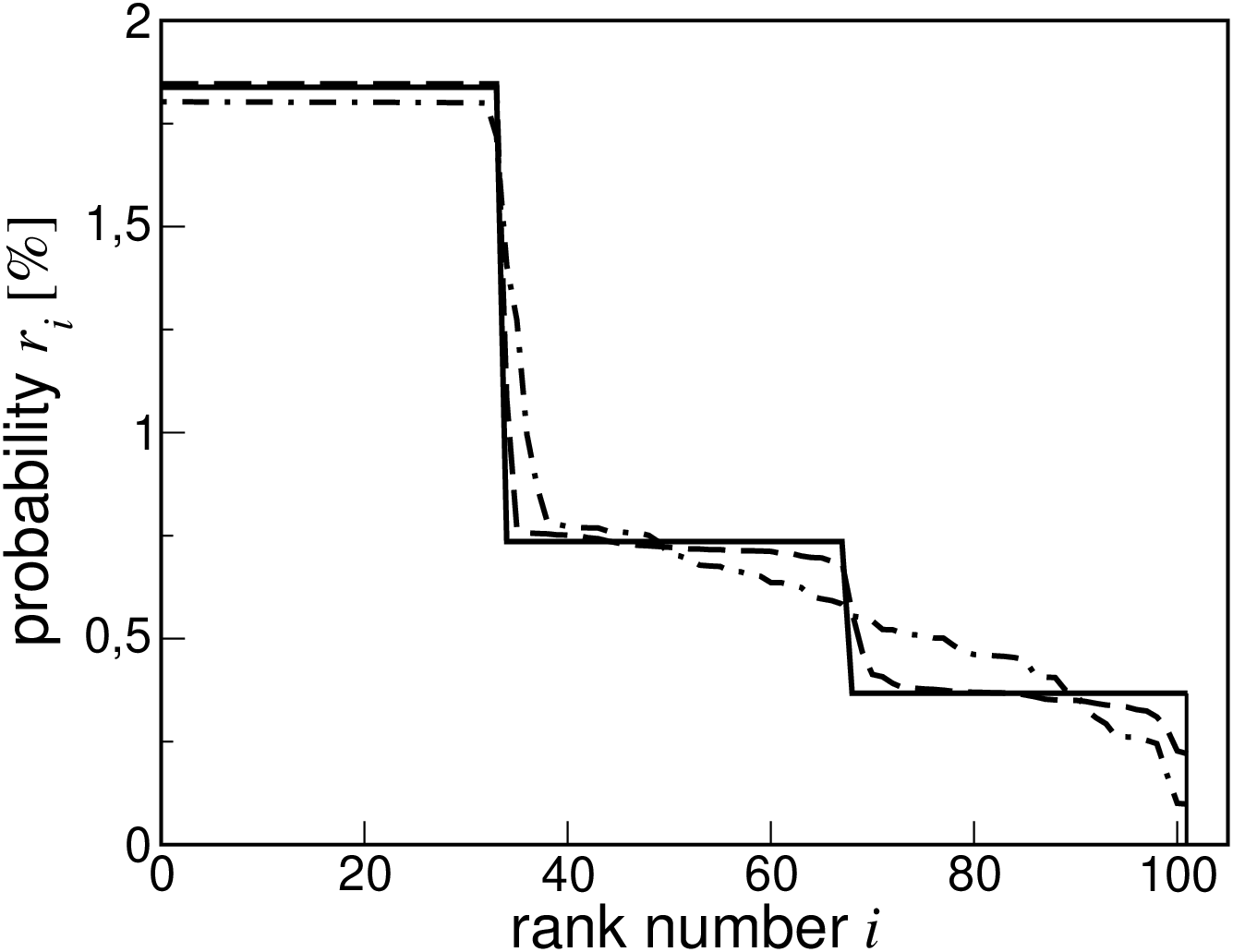}}
  \caption{Top, solid lines: rank-ordered normalized frequencies 
    $r_i^{\rm exp}$ generated from the step-wise probability
    distribution Eq. \eqref{eq:unfairdie} (data scaled by $M$). The
    upper curve corresponds to the sample size $M=5000$, the lower one
    to $M=2000$. These curves serve as input to our algorithm. Top,
    dashed lines: corresponding expected observed probabilities
    $r_i^*$ for sample sizes $M=5000$ and $M=2000$, respectively, as
    generated from the optimized set of probabilities. Bottom: solid
    line: original probability distribution given by Eq.
    \eqref{eq:unfairdie}. Dashed curve: rank-ordered probabilities as
    inferred by the described optimization algorithm based on a sample
    of size $M=5000$. Dot-dashed curve: same but for $M=2000$.}
  \label{fig:guess}
\end{figure}

We wish to demonstrate the performance of this algorithm by an
example. Using the step-wise probability distribution given by Eq.
\eqref{eq:unfairdie} we have drawn two samples of sample sizes
$M=2000$ and $M=5000$, respectively. The according rank-ordered
frequencies $r_i^{\rm exp}M$ are shown in Fig. \ref{fig:guess} (upper
plot, solid lines). These values serve as input (experimental data) to
our algorithm, i.e., we apply the algorithm to re-infer the true
step-wise probability distribution from these samples.
Applying the algorithm to these data we obtain the results shown in
the lower part of Fig. \ref{fig:guess}. Using the larger data set
$M=5000$ (dashed line) we reproduced the original function (solid
line) up to a good accuracy. Given the significant deformation of the
measured frequencies shown in the upper part of the figure, the
quality of the result surprises. Even for $M=2000$ where in the upper
part of the figure (lower dashed line) the three-step function can
hardly be recognized, the agreement of the numerical result of the
optimization procedure (lower plot, dot-dashed line) and the original
set of probabilities (solid line) is agreeable.

\begin{figure}[htbp]
  \centerline{\includegraphics[width=7cm,clip]{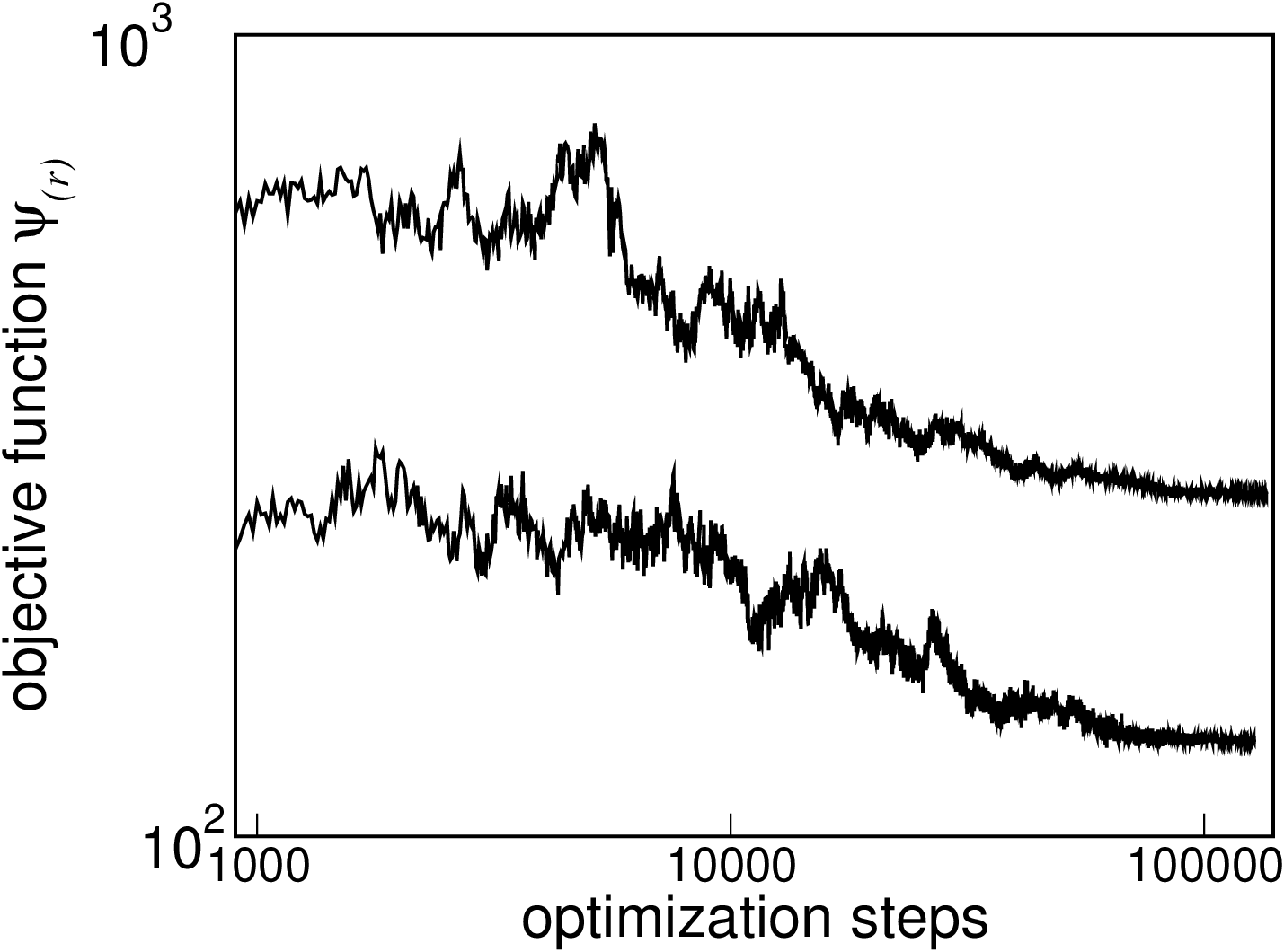}}
  \caption{Deviation of the input frequency distributions from 
    the frequency distributions which have been generated from the
    optimized probability distribution as defined by Eq.
    \eqref{eq:deviat} over the number of iteration cycles. The upper
    line shows the deviation for $M=5000$, the lower for $M=2000$.}
  \label{fig:guesserr}
\end{figure}

Figure \ref{fig:guesserr} shows the deviation of the input frequency
distributions from the frequency distributions which have been
generated from the optimized probability distribution, according to
Eq. \eqref{eq:deviat}.
After about 100,000 optimization loops the result does not improve
anymore. We expect that at this level the accuracy of the
approximation $k_i^{\rm exp}\approx \left<k_i\right>$ is reached.

\section{Discussion and perspectives}

In this article we propose a method to reconstruct the true
probability distribution of species from a set of frequency
distributions obtained from a small sample. This method relies on the
fact that the observed rank-ordered distribution of probabilities
$r^*_i$ for a finite sample of size $M$ can be predicted from the true
rank-ordered distribution $r_i$. Although we have not given any
mathematical strict proof of the theoretical expression of $r^*_i$, we
have shown that this method gives quantitatively good results for the
cases studied.

As mentioned, the rank-ordered distribution contains less information
than the species-ordered distribution: the identity of the species is
lost. Nevertheless, many statistical quantities are invariant with
respect to the species labeling. Any quantity defined as the sum over
all species of a function of their probability is insensitive to the
order of the species. The Shannon entropy, for example, can be written
as
\begin{equation}
H = - \sum_{i, \rm all species} p_i \log p_i = 
- \sum_{i, \rm all ranks} r_i \log r_i\,.
\end{equation}

Equation \eqref{eq:conc} allows for the prediction of the value of the
{\em observed entropy}, defined by
\begin{equation}
H^{*}=-  \sum_{i=1}^{N^*} r^*_i \log r^*_i = - \sum_{i=1}^{M} k_i^{\rm exp} \frac{i}{M}
\log{\frac{i}{M}}\,. 
\end{equation}
This quantity is experimentally accessible and serves in practical
applications as a measure for deviations from the equidistribution.
Since even for a perfect equidistribution of the species $H^*$
deviates from the entropy $H=\log{N}$ due to finite size effects as
shown in Sec. \ref{sec:rank} empirical entropies which are based on
different sample size $M$ cannot be compared directly with each other.
The method proposed here enables us to subdivide the deviations
$H^*-H$ into a part due to the finite sample size $M$,
\begin{equation}
- \sum_{i=1}^{M} \left<k_i\right> \frac{i}{M} \log{\frac{i}{M}}\,, 
\label{eq:expr}
\end{equation}
where the $\left<k_i\right>$ are given by Eq. \eqref{eq:Mom}, and a
part which is related to the true deviations from the equiprobability
distribution. This way we can compare also distributions which are
based on different sample sizes.  In the same way we can also quantify
sample-size independent deviations from any other distribution if we
compute the expected cluster sizes in the expression \eqref{eq:expr}
due to Eq. \eqref{eq:binomExpUnfair}.

Another question which can be solved using the methods developed here
is the evaluation of mean values of fluctuating quantities
\begin{equation}
  \label{eq:a}
\left<A\right> = \sum_{i=1}^N A_i p_i \,.
\end{equation}
This quantity may be determined by the following procedure: We
introduce first the set $a_i = A_i p_i$ and the corresponding observed
set
\begin{equation}
  \label{eq:b}
a_i^* = \frac{a_i M_i}{M} \,. 
\end{equation}
The set of the rank ordered numbers $a_i$ and, therefore, $\left<
  A\right>$ may be determined from the observed quantities $a_i^*$ in
precisely the same way as shown for the probabilities in this paper.
We believe that this new method to estimate mean values may have many
interesting applications.

The described correction algorithms have been applied to some
biological relevant examples such as the spatial distribution of point
mutations in genes \cite{PFG}.

\section*{Acknowledgment}
The authors are grateful to Jan Freund for discussion.


\end{document}